\documentclass[a4paper,11pt]{article}

\usepackage{natbib}

\usepackage{amssymb,type1cm}

\usepackage[dvipdfm,bookmarks=true,bookmarksopen=true]{hyperref}

\title{Computability of simple games:  A characterization and application to the core\thanks{%
Journal of Mathematical Economics (2007), 
\href{http://dx.doi.org/10.1016/j.jmateco.2007.05.012}{doi:10.1016/j.jmateco.2007.05.012}.
This pre-print contains Appendix~\ref{not-in-JME}, which is not contained in the JME article.
The proof of Theorem~4 of the JME article contains paragraphs that are incorrectly indented,
which makes it hard to grasp the scope of each step of the proof.
This pre-print is free of that flaw.}}

\author{Masahiro Kumabe \\
Kanagawa Study Center, The University of the Air\\
 2-31-1 Ooka, Minami-ku, Yokohama
232-0061, Japan 
\and 
H.  Reiju Mihara\thanks{Corresponding author. \protect\\
\emph{URL:} \url{http://econpapers.repec.org/RAS/pmi193.htm} (H.R. Mihara).}\\
Graduate School of Management, Kagawa University\\
Takamatsu 760-8523, Japan}

\date{August 2007}

\newcommand{\B}{\mathcal{B}}
\newcommand{\F}{\mathbf{F}}
\newcommand{\N}{\mathbb{N}}
\newcommand{\REC}{\mathrm{REC}}
\newcommand{\CRec}{\mathrm{CRec}}

\newcommand{\A}{\mathcal{A}}
\newcommand{\p}{\mathbf{p}}
\newcommand{\pref}{\succ_i^\p}
\newcommand{\profile}{\mbox{$(\pref)_{i\in N'}$}}

\newcommand{\xprefy}{\{\,i\in N':{x\pref y}\,\}}
\newcommand{\profs}{\A^{N'}_\B}

\newcommand{\qed}{\enspace\enspace \vrule height 6pt width5pt
depth2pt}

\usepackage{theorem}

\newtheorem{theorem}{Theorem}
\newtheorem{prop}[theorem]{Proposition}
\newtheorem{cor}[theorem]{Corollary}
\newtheorem{lemma}[theorem]{Lemma}

{\theorembodyfont{\normalfont}
\newtheorem{definition}[theorem]{Definition}
\newtheorem{example}[theorem]{Example}
\newtheorem{remark}[theorem]{Remark}
}

\newenvironment{proof}{\emph{Proof}.}{\qed\bigskip}
\newenvironment{subproof}{\emph{Proof}.}{\enspace$\|$\medskip}

\begin{document} 

\maketitle 


\begin{abstract} 
The class of algorithmically computable simple games 
(i) includes the class of games that have finite carriers and (ii) is included in the class
of games that have finite winning coalitions.
This paper characterizes computable games,
strengthens the earlier result that computable games violate anonymity,
and gives examples showing that the above inclusions are strict.
It also extends Nakamura's theorem about the nonemptyness of the core
and shows that computable games have a finite Nakamura number, 
implying that the number of alternatives that the players can deal with rationally is restricted.


\emph{Journal of Economic Literature} Classifications:  C71, D71, C69, D90.

\end{abstract}

\pagebreak

\section{Introduction}

We investigate algorithmic computability of a particular class of coalitional games (cooperative games), 
called \emph{simple games} (voting games).  
One can think of simple games as representing voting methods; alternatively,
as representing ``manuals'' or ``contracts.'' 
We give a characterization of computable simple games; it implies that a computable simple game uses
information about only finitely many players, but how much information it uses depends on each coalition.
We also apply the characterization to the theory of the core.
For the latter application, we extend Nakamura's 
theorem (\citeyear{nakamura79}) regarding the core of simple games to the framework
where not all subsets of players are deemed to be a coalition.

\subsection{Computability analysis of social choice}

Most of the paper (except the part on the theory of the core) 
can be viewed as a contribution to the foundations of 
\emph{computability analysis of social choice}, 
which studies algorithmic properties of social decision-making.  
This literature includes \citet{kelly88}, 
\citet{lewis88}, \citet{bartholdi-tt89vs,bartholdi-tt89cd},
and \citet{mihara97et,mihara99jme,mihara04mss},
who study issues in social choice using \emph{recursion theory} 
(the theory of computability and complexity).\footnote{
These works, which are mainly concerned with the complexity of rules or cooperative games in themselves,
can be distinguished from the closely related studies of the complexity of \emph{solutions} for cooperative games, 
such as \citet{deng-p94} and \citet{fang-zcd02}; %
they are also distinguished from the studies, such as \citet{takamiya-t0701}, of the complexity of deciding
whether a given cooperative game has a certain property.
(More generally, applications of recursion 
theory to economic theory and game theory include
\citet{spear89}, \citet{canning92}, \citet{anderlini-f94},
\citet{anderlini-s95},  \citet{prasad97},
\citet{richter-w99jme, richter-w99et}, and \citet{evans-t01}.
See also \citet{lipman95} and \citet{rubinstein98} for surveys of the literature on 
bounded rationality in these areas.)}

The importance of computability in social choice theory would be unarguable.
First, the use of the language by social choice theorists suggests the importance.
For example, Arrow defined a social welfare function to be 
a ``\emph{process or rule}'' which, for each profile of 
individual preferences, ``states'' a corresponding social 
preference \citep[p.~23]{arrow63}, and called the function a 
``\emph{procedure}'' \citep[p.~2]{arrow63}.  Indeed, he later 
wrote \citep[p.~S398]{arrow86} in a slightly different context, ``The next 
step in analysis, I would conjecture, is a more consistent assumption of 
\emph{computability} in the formulation of economic hypotheses'' (emphasis added).
Second, there is a normative reason.  Algorithmic 
social choice rules specify the procedures in such a way that the same 
results are obtained irrespective of who carries out a computation, leaving 
no room for personal judgments.  In this sense, 
computability of social choice rules formalizes the notion of ``due 
process.''\footnote{
\citet{richter-w99jme} give reasons for studying 
computability-based economic theories from the viewpoints of 
bounded rationality, computational economics, and complexity analysis.
These  reasons partially apply to studying computable rules in social choice.}

\subsection{Simple games with countably many players}\label{Intro2}

Simple games have been central to the study of social choice 
\citep[e.g.,][]{banks95,austensmith-b99,peleg02hbscw}.
\emph{Simple games}  on an algebra of coalitions of players assign either 0 or~1 to each coalition
(member of the algebra).
In the setting of players who face a yes/no question, 
a coalition intuitively describes those players who vote yes.
A simple game is characterized by its \emph{winning} coalitions---those assigned the value~1.
(The other coalitions are \emph{losing}.)
Winning coalitions are understood to be those coalitions whose unanimous votes are decisive.


When there are only \emph{finitely} many players, we can construct a finite table listing 
all winning coalitions.  Computability is automatically satisfied, since such a table gives an 
algorithm for computing the game.
The same argument does not hold when there are \emph{infinitely} many players.
Indeed, some simple games are noncomputable,
since there are uncountably many simple games but only countably many computable ones
 (because each computable game is associated with an algorithm).
 
Taking the ``fixed population'' approach,\footnote{
There are two typical approaches to introducing infinite population to a social choice model.
In the ``variable population'' approach, 
players are potentially infinite, but each problem (or society) involves only finitely many
players.  Indeed, well-known schemes, such as simple majority rule, 
unanimity rule, and the Condorcet and the Borda rules, are all algorithms that 
apply to problems of any finite size.
\citet{kelly88} adopts this approach, giving examples of noncomputable social choice rules.
In the ``fixed population'' approach, which we adopt, each problem involves the whole set of 
infinitely many players.  This approach dates back to \citet{downs57}, who
consider continuous voter distributions.
The paper by \citet{banks-dl06} is a recent example of this approach to political theory.}  %
we consider a fixed infinite set of players in this study of simple games.
Roughly speaking, a simple game is \emph{computable} if there is a Turing program (finite algorithm) that can decide
from a description (by integer) of each coalition whether it is winning or losing.
To be more precise, we have to be more specific about what we mean by a ``description'' of a coalition.
This suggests the following:
First, since each member of a coalition should be describable in words (in English), it is natural to assume that the
set $N$ of (the names of) players is countable, say, $N=\N=\{0,1,2, \dots \}$.
Second, since one can describe only countably many coalitions, we have to restrict coalitions.
Finite or cofinite coalitions can be described by listing their members or nonmembers completely.
But restricting coalitions to these excludes too many coalitions of interest---such as the set of even numbers.
A natural solution is to describe coalitions by a Turing program that can decide for the name of each player whether
she is in the coalition.  Since each Turing program has its code number (G\"{o}del number),
the coalitions describable in this manner are describable by an integer, as desired.
Our notion of computability (\emph{$\delta$-computability}) focuses on this
class of coalitions---\emph{recursive} coalitions---as well as the method (characteristic index) of describing them.

A fixed population of countably many \emph{players} arises not only in voting but in other contexts, 
such as a special class of multi-criterion decision making---depending on how we interpret a ``player'':
\begin{description}
\item [Simulating future generations] 
One may consider countably many \emph{players} (people) extending into the indefinite future.

\item [Uncertainty]
One may consider finitely many \emph{persons} facing countably many states of the world~\citep{mihara97et}:
each \emph{player} can be interpreted as a particular \emph{person} in a particular \emph{state}.
The decision has to be made before a state is realized and identified.
(This idea is formalized by \citet{gomberg-mt05}, who introduce
``$n$-period coalition space,'' where $n$ is the number of persons.)

\item [Team management]
Putting the right people (and equipment) in the right places is basic to team management.\footnote{In line with much of 
cooperative game theory, we put aside the important problems of economics of organization, such as coordinating the
activities of the team members by giving the right incentives.}
To ensure ``due process'' (which is sometimes called for), can a manager of a company write 
a ``manual'' (computable simple game\footnote{Like \citet{anderlini-f94}, who view contracts as algorithms, we view ``manuals'' as algorithms.  They derive contract incompleteness through computability analysis.}) elaborating the conditions that a team must meet?

Fix a particular task such as operating an exclusive agency of the company.%
\footnote{Extension to finitely or countably many tasks is straightforward.
Redefine a \emph{team} as consisting of members, equipment, \emph{and} tasks.
Then introduce a player for each task.  
Since a task can be regarded as a negative input, it will be more natural to assign
$0$ to those tasks undertaken and $1$ to those not undertaken (think of the monotonicity condition).}
A \emph{team} consists of members (people) and equipment.
The manager's job is to organize or give a licence to a team that satisfactorily performs the task.
Each member (or equipment) is described by attributes such as skills, position, availability at a particular time and place (in case of equipment such as a computer, the attributes may be the kind of operating system, the combination of software that may run at the same time, as well as hardware and network specifications).
Each such \emph{attribute} can be thought of as a particular yes/no question, 
and there are countably many such questions.\footnote{According to a certain approach \citep[e.g.,][]{gilboa90}
to modeling scientific inquiry, 
a ``state'' is an infinite sequence of $0$'s and $1$'s (answers to countably many questions) 
and a ``theory'' is a Turing program describing a state.
This team management example is inspired by this approach in philosophy of science.
If we go beyond the realm of social choice, we can indeed find many other interpretations
having a structure similar to this example, such as elaborating the conditions for a certain medicine to take the 
desired effects and deciding whether a certain act is legal or not.}


Here, each \emph{player} can be interpreted as a particular attribute of a particular member (or equipment).%
\footnote{From the viewpoint of ``due process,'' it would be reasonable to define a simple game not for the 
set of (the names of) members but for the set of attributes.  
(This is particularly important where games cannot meet anonymity.) 
Considering characteristic games for the set of attributes 
(``skills'') can make it easy to express certain allocation problems and give solutions to them 
\citep[see][]{yokoo-05aaai}.}
In other words, each \emph{coalition} is identified with a $0$-$1$ ``matrix'' of 
finitely many rows (each row specifying a member) and 
countably many columns (each column specifying a particular attribute).\footnote{Since different 
questions may be interrelated,
some ``matrix'' may not make much sense.  One might thus want to restrict admissible ``matrices.''  
This point is not crucial to our discussion, provided that there are infinitely many 
admissible ``matrices'' consisting of infinitely many $0$'s and infinitely many $1$'s (in such cases
characteristic indices are the only reasonable way of naming coalitions).}

\end{description}

\subsection{Overview of the results}


Adopting the above notion of computability for simple games, 
\citet{mihara04mss} gives a sufficient condition and necessary conditions 
for computability.  The sufficient condition \citep[Proposition~5]{mihara04mss} is intuitively plausible: 
simple games with a \emph{finite carrier} (such games are in effect finite, ignoring all except finitely many,
fixed players' votes) are computable.
A necessary condition \citep[Corollary~10]{mihara04mss} in the paper seems to exclude ``nice'' 
(in the voting context) infinite games:
computable simple games have both finite winning coalitions and cofinite losing coalitions.
He leaves open the questions (i)~whether there exists a computable simple game that has no finite carrier and
(ii)~whether there exists a noncomputable simple game that has both finite winning coalitions and cofinite 
losing coalitions. 
The first of these questions is particularly important since if the answer were no, then
only the games that are in effect finite would be computable, a rather uninteresting result.
The answers to these questions (i) and~(ii) are affirmative.  
We construct examples in Section~\ref{examples} to show their existence.
The construction of these examples depends in essential ways on Proposition~\ref{cutprop}
(which gives a necessary condition for a simple game to be computable) or on
the easier direction of Theorem~\ref{delta0det} (which gives a sufficient condition).
In contrast, the results in \citet{mihara04mss} are not useful enough for 
us to construct such examples.


Theorem~\ref{delta0det} gives a necessary and sufficient condition for simple games to be computable.
The condition roughly states that ``finitely many, unnecessarily fixed players matter.''  

To explain the condition, let us introduce the notion of a ``determining string.''   
Given a coalition~$S$, its \emph{$k$-initial segment} is the string of $0$'s and $1$'s of length $k$
whose $j$th element (counting from zero) is $1$ if $j\in S$ and is $0$~if $j\notin S$.  
For example, if $S=\{0,2,4\}$, 
its $0$-initial segment, $1$-initial segment, \ldots, $8$-initial segment, \ldots are, respectively, the empty string,
the string~$1$, the string $10$, the string~$101$, the string $1010$, the string $10101$,
the string $101010$, the string $1010100$, the string $10101000$, \ldots.
We say that a (finite) string $\tau$ is \emph{winning determining} if any coalition~$G$ extending $\tau$
(i.e., $\tau$ is an initial segment of~$G$) is winning.  We define \emph{losing determining} strings similarly.

The necessary and sufficient condition for computability according to Theorem~\ref{delta0det}
is the following: there are computably listable sets 
$T_0$ of losing determining strings and $T_1$ of winning determining strings such that 
any coalition has an initial segment in one of these sets.  
In the above example, the condition 
implies that at least one string from among the empty string, $1$, $10$, \ldots, $10101000$, 
\ldots is in $T_0$ or in $T_1$---say, $1010$ is in $T_1$.
Then any coalition of which $0$ and $2$ are members but $1$ or $3$ is not, is winning.
In this sense, \emph{one can determine whether a coalition is winning or losing
by examining only finitely many players' membership}.  
In general, however, \emph{one cannot do so by picking finitely many players before a coalition is given}.

Theorem~\ref{delta0det} has an interesting implication for the nature of  ``manuals'' or ``contracts,'' 
if we regard them as being composed of computable simple games
 (e.g., the team management example in Section~\ref{Intro2}).
Consider how many ``criteria''  (players; e.g., member-attribute pairs) are needed for a ``manual'' to determine 
whether a given ``situation'' (coalition; e.g., team) is ``acceptable'' (winning; e.g., satisfactorily performs a given task).
\emph{While increasingly complex situations may require increasingly many criteria, 
no situation (however complex) requires infinitely many criteria.}
The conditions  (such as ``infinitely many of the prime-numbered criteria must be met'') 
based on infinitely many criteria are ruled out.

The proof of Theorem~\ref{delta0det} uses the \emph{recursion theorem}.
It involves much more intricate arguments of recursion theory
than those in \citet{mihara04mss} giving only a partial characterization of the computable 
games.\footnote{Theorem~\ref{delta0det} can also be derived from results in \citet{kreisel-ls59}
and \citet{ceitin59}.  See Remark~\ref{kreisel59}.}

A natural characterization result might relate computability to well-known properties of simple games, 
such as monotonicity, properness, strongness, and nonweakness.
Unfortunately, we are not likely to obtain such a result: as we clarify in a companion paper~\citep{kumabe-m06csg64}, 
computability is ``unrelated to'' the four properties just mentioned.


The earlier results~\citep{mihara04mss} are easily obtained from Theorem~\ref{delta0det}.
For example, if a computable game has a winning coalition, then,
an initial segment of that coalition is winning determining, implying that (Proposition~\ref{d0neg1cor}) 
the game has a finite winning coalition and a cofinite winning coalition.
We give simple proofs to some of these results in Section~\ref{applications}.
In particular, Proposition~\ref{nonfinanonymous} strengthens the earlier result \citep[Corollary~12]{mihara04mss} 
that computable games violate anonymity.\footnote{
Detailed studies of anonymous rules based on infinite simple games include
\citet{mihara97scw}, \citet{fey04}, and \citet{gomberg-mt05}.}

\subsection{Application to the theory of the core}

Most cooperative game theorists are more interested in the properties of a \emph{solution} (or value) 
for games than in the properties of a game itself.
In this sense, Section~\ref{core} deals with more interesting applications of Theorem~\ref{delta0det}.
(Most of the section is of independent interest, and can be read without a knowledge of recursion theory.)

Theorem~\ref{nakamura-thm} is our main contribution to the study of 
acyclic preference aggregation rules in the spirit of 
Nakamura's theorem (\citeyear{nakamura79}) on the core of simple games.\footnote{
\citet{banks95}, \citet{truchon95}, and \citet{andjiga-m00} are recent contributions
to this literature.
(Earlier papers on acyclic rules can be found in \citet{truchon95} 
and \citet{austensmith-b99}.)
Most works in this literature (including those just mentioned) consider finite sets of players.
\citet{nakamura79} considers arbitrary (possibly infinite) sets of players and the algebra of all subsets of players.
In contrast, we consider arbitrary sets of players and \emph{arbitrary algebras} of coalitions.}

Combining a simple game with a set of alternatives and a profile of individual preferences, we define a 
\emph{simple game with (ordinal) preferences}.
Nakamura's theorem (\citeyear{nakamura79}) gives a necessary and sufficient condition for a simple game with
preferences to have a nonempty core for all profiles: 
the number of alternatives is below a certain number, called the \emph{Nakamura number} of the simple game.
We extend (Theorem~\ref{nakamura-thm}) Nakamura's theorem to the framework 
where simple games are defined on an arbitrary algebra of coalitions
(so that not all subsets of players are coalitions).
\emph{It turns out that our proof for the generalized result is more elementary than Nakamura's original proof;
the latter is more complex than need be.}

Since computable (nonweak) simple games have a finite winning coalition, 
we can easily prove that they have a finite Nakamura number (Corollary~\ref{nakamura-finite}).
Theorem~\ref{nakamura-thm} in turn implies (Corollary~\ref{core-nakamura}) that 
if a game is computable, 
the number of alternatives that the set of players can deal with rationally is restricted by this number.
We conclude Section~\ref{core} with Proposition~\ref{filter-nakamura}, which suggests the fundamental 
difficulty of obtaining computable aggregation rules in Arrow's setting (\citeyear{arrow63}),
even after relaxing the transitivity requirement for (weak) social preferences.\footnote{
\citet{mihara97et,mihara99jme} studies computable aggregation rules \emph{without} relaxing the
transitivity requirement; these papers build on \citet{armstrong80,armstrong85}, who generalizes
\citet{kirman-s72}.}

\section{Framework}

\subsection{Simple games}\label{notions}

Let $N=\N=\{0,1,2, \dots \}$ be a countable set of (the names of) 
players.  Any \textbf{recursive} (algorithmically decidable) 
subset of~$N$ is called a \textbf{(recursive) coalition}.

Intuitively, a simple game describes in a crude manner the power 
distribution among \emph{observable} (or describable) subsets of players.  Since the cognitive 
ability of a human (or machine) is limited, it is not natural to 
assume that all subsets of players are observable when there are 
infinitely many players.  We therefore assume that only 
\textbf{recursive} subsets are 
observable.  This is a natural assumption in the present context, 
where algorithmic properties of simple games are investigated.  
According to \emph{Church's thesis} \citep[see][]{soare87,odifreddi92}, the recursive coalitions 
are the sets of players for which there is an algorithm that can decide for 
the name of each player whether she is in the set.\footnote{\citet{soare87} and \citet{odifreddi92}
give a more precise definition of \emph{recursive sets} as well as detailed discussion of recursion theory.
\citet{mihara97et,mihara99jme} contain short reviews of recursion theory.}
Note that \textbf{the class~$\REC$ of recursive 
coalitions} forms a \textbf{Boolean algebra}; that is, it includes $N$ 
and is closed under union, intersection, and complementation.
(We assume that observable coalitions are recursive, not just r.e.\ (\emph{recursively enumerable}).  
\citet[Remarks~1 and 16]{mihara04mss} gives three reasons: 
nonrecursive r.e.\ sets are observable in a very limited sense; 
the r.e.\  sets do not form a Boolean algebra;
no satisfactory notion of computability can be defined if a simple game is 
defined on the domain of all r.e.\ sets.)

Formally, a \textbf{(simple) game} is a collection~$\omega\subseteq\REC$ of (recursive) coalitions.
We will be explicit when we require that $N\in \omega$.
The coalitions in $\omega$ are said to be \textbf{winning}.  
The coalitions not in $\omega$ are said to be \textbf{losing}. 
One can regard a simple game as a function from~REC to $\{0,1\}$, assigning the value 1 or 0 to each 
coalition, depending on whether it is winning or losing.

We introduce from the theory of cooperative games a few basic 
notions of simple games~\citep{peleg02hbscw,weber94}.\footnote{The desirability of these properties depends, 
of course, on the context.  Consider the team management example in Section~\ref{Intro2}, for example.
Monotonicity makes sense, but may be too optimistic (adding a member may turn an acceptable team into an
unacceptable one).  Properness may be irrelevant or even undesirable 
(ensuring that a given task can be performed by 
two non-overlapping teams may be important from the viewpoint of reliability).
This observation does not diminish the contribution of the main theorem (Theorem~\ref{delta0det}),
which does not refer to these properties.
In fact, one can show~\citep{kumabe-m06csg64}
 that computability is ``unrelated to'' monotonicity, properness, strongness, and weakness.}
A simple game $\omega$ is said to be 
\textbf{monotonic} if for all coalitions $S$ and $T$, the 
conditions $S\in \omega$ and $T\supseteq S$ imply $T\in\omega$.  
$\omega$ is \textbf{proper} if for all recursive coalitions~$S$, 
$S\in\omega$ implies $S^c:=N\setminus S\notin\omega$.  $\omega$ is 
\textbf{strong} if for all coalitions~$S$, $S\notin\omega$ 
implies $S^c\in\omega$.  $\omega$ is \textbf{weak} if 
$\omega=\emptyset$ or
the intersection~$\bigcap\omega=\bigcap_{S\in\omega}S$ of the winning coalitions is nonempty.  
The members of $\bigcap\omega$ are called \textbf{veto players}; they 
are the players that belong to all winning coalitions.  
(The set $\bigcap\omega$ of veto players may or may not be observable.)
$\omega$ is \textbf{dictatorial} if there exists some~$i_0$ (called a 
\textbf{dictator}) in~$N$ such that $\omega=\{\,S\in\REC: i_0\in 
S\,\}$.  Note that a dictator is a veto player, but a veto player is 
not necessarily a dictator.

We say that a simple game $\omega$ is \textbf{finitely anonymous} if for any finite permutation $\pi: N\to N$ 
(which permutes only finitely many players) and for 
any coalition $S$, we have 
$S\in \omega \iff \pi(S) \in \omega$.  
In particular, finitely anonymous games treat any two coalitions with the same finite number of players equally.
Finite anonymity is a notion much weaker than 
the version of anonymity that allows any (measurable) permutation $\pi: N\to N$.
For example, free ultrafilters (nondictatorial ultrafilters) defined below are finitely anonymous.  

A \textbf{carrier} of a simple game~$\omega$ is a coalition $S\subset N$ 
such that
\[ 
T\in\omega \iff S\cap T\in \omega
\]
for all coalitions~$T$.
We observe that if $S$ is a carrier, then so is any coalition $S'\supseteq S$.

Finally, we introduce a few notions from the theory of Boolean 
algebras~\citep{koppelberg89}; they can be regarded as properties of 
simple games.  
A monotonic simple game~$\omega$ satisfying
$N\in \omega$ and $\emptyset\notin\omega$
is called a \textbf{prefilter} if it has the finite intersection property:
if $\omega'\subseteq \omega$ is finite, then $\bigcap\omega'\neq \emptyset$.
Intuitively, a prefilter consists of ``large'' coalitions.
A prefilter is \textbf{free} if and only if it is nonweak (i.e., it has no veto players).
A free prefilter does not contain any finite coalitions (Lemma~\ref{nakamura-ceiling}).
A prefilter~$\omega$ is a \textbf{filter} if it is closed with respect to finite intersection: 
if $S$, $S'\in\omega$, then $S\cap S'\in \omega$.
The \textbf{principal filter generated by $S$} is
$\omega=\{T\in \REC: S\subseteq T\}$.  It is a typical example of a filter that is
not free; it has a carrier, namely, $S$.  A filter is a \textbf{principal filter} if it is 
the principal filter generated by some~$S$.
A filter~$\omega$ is called an \textbf{ultrafilter} if it is a 
strong simple game.  If $\omega$ is an ultrafilter, then $S\cup 
S'\in\omega$ implies that $S\in\omega$ or $S'\in\omega$.
An ultrafilter is free if and only if it is not dictatorial.

\subsection{An indicator for simple games}\label{indicators}

To define the notion of computability for simple games, we introduce below 
an indicator for them.  In order to do that, 
we first represent each recursive coalition by a characteristic index ($\Delta_0$-index).
Here, a number $e$~is a \textbf{characteristic index} for a coalition~$S$
if $\varphi_e$ (the partial function computed by the Turing program with code number~$e$)
is the characteristic function for~$S$. 
Intuitively, a characteristic index for a coalition describes
the coalition by a Turing program that can decide its membership.
The indicator then assigns the value 0 or 1 to each 
number representing a coalition, depending on whether the 
coalition is winning or losing.  When a number does not represent a 
recursive coalition, the value is undefined.

Given a simple game $\omega$, its \textbf{$\delta$-indicator} is the partial 
function~$\delta_\omega$ on~$\N$ defined by
\begin{equation}
	\label{d:eq}
	\delta_\omega(e)=\left\{
	    \begin{array}{ll}
		1 & \mbox{if $e$ is a characteristic index for a recursive
	set in $\omega$},  \\
		0 & \mbox{if $e$ is a characteristic index for a recursive
	set not in $\omega$},  \\
		\uparrow & \mbox{if $e$ is not a characteristic
	index for any recursive set}.
	\end{array}
	\right.
\end{equation}
Note that $\delta_\omega$ is well-defined since each $e\in\N$ can be a 
characteristic index for at most one set.

\subsection{The computability notion}
\label{comp:notions}

We now introduce the notion of \emph{$\delta$-computable} simple games.
We start by giving a scenario or intuition underlying the notion of $\delta$-computability.
A number (characteristic index) representing a coalition
(equivalently, a Turing program that can decide the membership of a coalition)
is presented by an inquirer to the aggregator (planner), 
who will compute whether the coalition is winning or not.
Though there is no effective (algorithmic) procedure to decide whether a number given by the 
inquirer is legitimate (i.e., represents some recursive coalition),
a human can often check manually (non-algorithmically) if such a number is a legitimate representation.
We assume that the inquirer gives the aggregator only those indices that he has checked and proved its legitimacy.
This assumption is justified if we assume that the aggregator always demands such proofs.
The aggregator, however, cannot know a priori which indices will possibly be presented to her.
(There are, of course, indices unlikely to be used by humans.  
But the aggregator cannot a priori rule out some of the indices.)
So, \emph{the aggregator should be ready to compute whenever a legitimate representation
 is presented to her}.\footnote{
An alternative notion of computability might use a ``multiple-choice format,''
 in which the aggregator gives possible indices that the inquirer can choose from. 
Unfortunately, such a ``multiple-choice format'' would not work as one might wish
\citep[Appendix~A.1]{kumabe-m07csgcc}.}  
This intuition justifies the following condition of computability.\footnote{
\citet{mihara04mss} also proposes a stronger condition, \emph{$\sigma$-computability}.
We discard that condition since it is too strong a notion of computability (Proposition~3 of that paper;
for example, even \emph{dictatorial} games are not $\sigma$-computable).} %

\begin{description} \item[$\delta$-computability] $\delta_\omega$ has 
an extension to a partial recursive function.  
\end{description}

Instead of, say, $\delta$-computability, one might want to require the indicator~$\delta_\omega$ \emph{itself} 
(or its extension that gives a number different from $0$ or~$1$ whenever $\delta_\omega(e)$ is undefined)
to be partial recursive \citep[Appendix~A]{mihara04mss}.  
Such a condition cannot be satisfied, however, since the domain of~$\delta_\omega$ is
not r.e. \citep[Lemma~2]{mihara97et}.

\section{A Characterization Result}\label{mainresult}


\subsection{Determining strings}

The next lemma states that for any coalition~$S$ of a 
$\delta$-computable simple game,  
there is a cutting number~$k$ such 
that any \emph{finite} coalition~$G$ having the same $k$-players 
as~$S$ (that is, $G$ and $S$ are equal if players~$i\geq k$ are 
ignored) is winning (losing) if $S$ is winning (losing).  Note that 
if $k$ is such a cutting number, then so is any $k'$ greater than~$k$.

\medskip

\textbf{Notation}.  We identify a natural number~$k$ with the finite 
set $\{0,1,2,\ldots,k-1\}$, which is an initial segment of~$\N$.  
Given a coalition $S\subseteq N$, we write $S\cap k$ to represent the 
coalition $\{i\in S: i<k\}$ consisting of the members of $S$ whose 
name is less than~$k$.  
We call $S\cap k$ the \textbf{$k$-initial segment of $S$}, and view it 
either as a subset of~$\N$ or as the string $S[k]$ of length~$k$ of 0's and 1's 
(representing the restriction of its characteristic function to 
$\{0,1,2,\ldots,k-1\}$).  
For example, if $S=\{0, 2, 4\}$, we have 
$S[7]=1010100=\varphi_e(0)\varphi_e(1) \cdots \varphi_e(6)$,
where $e$ is a characteristic index for~$S$.
Note that if $G$ is a coalition and $G\cap 
k=S\cap k$ (that is, $G$ and $S$ are equal if players~$i\geq 
k$ are ignored), the characteristic function of~$G$ extends the $k$-initial 
segment (viewed as a string of 0's and 1's) of~$S$.

\begin{lemma} \label{cutlemma} 
Let $\omega$ be a $\delta$-computable simple game.  If $S\in\omega$, 
then there is an initial segment $k\geq 0$ of~$\N$ such that for any 
\emph{finite}~$G\in \REC$, if $G\cap k=S\cap k$, then 
$G\in\omega$.  Similarly, if $S\notin\omega$, then there is an initial 
segment $k\geq 0$ of~$\N$ such that for any \emph{finite}~$G\in 
\REC$, if $G\cap k=S\cap k$, then $G\notin\omega$.
\end{lemma}

\begin{proof} 
Let $S\in\omega$ and assume for a contradiction that there is no such 
initial segment~$k$.  Then, for each initial segment~$k$ of~$\N$, there 
is a finite coalition $G_k$ such that $G_k\cap k=S\cap k$ and 
$G_k\notin\omega$.  Note that we can find such $G_k$ recursively 
(algorithmically) in~$k$ since it is finite.

Let $K$ be a nonrecursive r.e.\ set such as $\{\,e:e\in W_e\,\}$.  
Since $K$ is r.e., there is a recursive set $R\subseteq\N\times\N$ 
such that $e\in K\Leftrightarrow \exists z R(e,z)$.  Define 
$g(e,u)=\mu y\leq u\ R(e,y)$ (i.e., the least $y\leq u$ such that 
$R(e,y)$) if such $y$ exists, and $g(e,u)=0$ otherwise.  Then $g$~is 
recursive.

Using the Parameter Theorem, define a recursive function~$f$ by\footnote{
Intuitively, given $e$, we define $\varphi_{f(e)}(u)$ at each step~$u$ as follows:
Let $u':=\mu y\ R(e,y)$ if $e\in K$;  $u':=+\infty$ otherwise.
Try to find $z \leq u$ such that $R(e, z)$. 
If $u$ is small (i.e., $u<u'$), there is no such $z$ (the first and the second cases);
for those~$u$,  $\varphi_{f(e)}$ takes the same value as the characteristic function for~$S$.
If $u$ is large enough (i.e., $u\geq u'=g(e,u)$, which is true only if $e\in K$),
there is such $z$ (the third and the fourth cases);
for those~$u$, $\varphi_{f(e)}$ takes the same value as the characteristic function for a certain
losing coalition~$G_k$ defined above (namely, $k=u'$).
Note that the construction of $G_k$ ensures that the characteristic functions for $S$ and 
for $G_k$ take the same values for small $u$ as well.} 
\begin{eqnarray*}
\varphi_{f(e)}(u)=1 & & \textrm{if $\neg\exists z\leq u\ R(e,z)$ and 
$u\in S$},\\
\varphi_{f(e)}(u)=0 & & \textrm{if $\neg\exists z\leq u\ R(e,z)$ and 
$u\not\in S$},\\
\varphi_{f(e)}(u)=1 & & \textrm{if $\exists z\leq u\ R(e,z)$ and $u\in 
G_{g(e,u)}$, and}\\
\varphi_{f(e)}(u)=0 & & \textrm{otherwise}.
\end{eqnarray*}

Now, on the one hand, $e\in K$ implies that $f(e)$ is a characteristic 
index for $G_{u'}\notin\omega$ for some~$u'$.  (\emph{Details}: Given 
$e\in K$, let $u'=\mu y\ R(e,y)$, which is well-defined since $\exists 
z R(e,z)$.  Then $\varphi_{f(e)}(u)=1$ iff (i)~$u<u'$ and $u\in S$ 
[that is, $u<u'$ and $u\in G_{u'}$] or (ii)~$u\geq u'$ and [since 
$g(e,u)=u'$ in this case] $u\in G_{g(e,u)}=G_{u'}$.  Thus 
$\varphi_{f(e)}(u)=1$ iff $u\in G_{u'}$.)  Hence 
$\delta_{\omega}(f(e))=0$.  On the other hand, $e\notin K$ implies 
that $f(e)$ is a characteristic index for $S\in\omega$.  Hence 
$\delta_{\omega}(f(e))=1$.

Since $\delta_{\omega}$ has an extension to a p.r.\ function (because 
$\omega$ is $\delta$-computable), the last paragraph implies that $K$ 
is recursive.  This is a contradiction.

\medskip

To prove the last half of the lemma, note that the set-theoretic 
difference $\hat{\omega}=\REC-\omega$ is also a 
$\delta$-computable simple game.  Let $S\notin\omega$.  Then the first 
half applies to $\hat{\omega}$ and $S\in\hat{\omega}$.  Since 
$G\in\hat{\omega}$ iff $G\notin\omega$, the desired result 
follows.\end{proof}

In fact, the coalition~$G$ in Lemma~\ref{cutlemma} need not be finite.  
Before stating an extension (Proposition~\ref{cutprop}) of Lemma~\ref{cutlemma}, we introduce
the notion of \emph{determining strings}:

\begin{definition}
Consider a simple game.  A string $\tau$ (of 0's and 1's) of 
length~$k\geq 0$ is said to be \textbf{determining} if either any 
coalition $G\in\REC$ extending $\tau$ (in the sense that $\tau$ is an 
initial segment of $G$, i.e., $G\cap k=\tau$) is winning or any 
coalition $G\in\REC$ extending $\tau$ is losing.  A string $\tau$ is 
said to be \textbf{determining for finite coalitions} if either any 
finite coalition~$G$ extending $\tau$ is winning or any finite 
coalition~$G$ extending $\tau$ is losing.  A string is  
\textbf{nondetermining} if it is not determining.
\end{definition}

Proposition~\ref{cutprop} below states that for $\delta$-computable simple 
games, (the characteristic function for) every coalition~$S$ has an 
initial segment $S\cap k$ that is determining.  (The number~$k-1$ may 
be greater than the greatest element, if any, of~$S$):

\begin{prop} \label{cutprop} 
Suppose that a $\delta$-computable simple game is given.  
\textup{(i)}~If a coalition $S$ is winning, then there is an initial segment $k\geq 0$ 
of~$\N$ such that for \emph{any} (finite or infinite) coalition~$G$, 
if $G\cap k=S\cap k$, then $G$ is winning.  
\textup{(ii)}~If $S$ is losing, then there is an initial segment $k\geq 0$ of~$\N$ such that for 
\emph{any} coalition~$G$, if $G\cap k=S\cap k$, then $G$ is losing.
\textup{(iii)}~If $S\cap k$ is an initial segment that is determining 
\emph{for finite coalitions}, then $S\cap k$ is an initial segment that is determining.
\end{prop}

\begin{proof}  
It suffices to prove~(i).  As a byproduct, we obtain~(iii).

Suppose $S\in\omega$, where 
$\omega$ is a $\delta$-computable simple game.  Then by the first half 
of Lemma~\ref{cutlemma}, there is $k\geq 0$ such that (a)~for any 
finite $G'$, if $G'\cap k=S\cap k$, then $G'\in\omega$.

To obtain a contradiction, suppose that there is $G\notin\omega$ such 
that (b)~$G\cap k=S\cap k$.  By the last half of Lemma~\ref{cutlemma}, 
there is $k'\geq 0$ such that (c)~for any finite $G'$, if $G'\cap 
k'=G\cap k'$, then $G'\notin\omega$.  Without loss of generality, 
assume $k'\geq k$.

Consider $G'=G\cap k'$, which is finite.  Then, on the one hand, since 
$G'\cap k'=G\cap k'$, we get $G'\notin\omega$ by (c).  On the other 
hand, since $k'\geq k$, we get $G'\cap k=G\cap k=S\cap k$ (the last 
equality by~(b)).  Then (a)~implies that $G'\in\omega$.  This is a 
contradiction.\end{proof}

\subsection{Characterization of computable games}

The next theorem characterizes $\delta$-computable simple games in 
terms of sets of determining strings.  Roughly speaking,
finitely many players determine whether a coalition is winning or losing.
Though we cannot tell in advance which finite set of players determines that, 
we can list such sets in an effective manner.

Note that $T_0\cup T_1$ in the theorem does not necessarily contain all determining strings.
(The $\Longrightarrow$ direction can actually be strengthened:  
we can find \emph{recursive}, not just r.e., sets $T_0$ and $T_1$ satisfying the conditions.
We do not prove the strengthened result, since we will not use it.)

\begin{theorem}\label{delta0det} 
A simple game~$\omega$ is $\delta$-computable if and only if there are 
an r.e.\ set~$T_0$ of losing determining strings and an r.e.\ 
set~$T_1$ of winning determining strings such that (the characteristic 
function for) any coalition has an initial segment in $T_0$ or in 
$T_1$.\end{theorem}

\begin{remark}\label{kreisel59}
We can derive Theorem~\ref{delta0det} from a result in \citet{kreisel-ls59}
and \citet{ceitin59}.  
In the working paper \citep[Appendix A.2]{kumabe-m07csgcc},  
following the terminology of \citet{odifreddi92},
we outline such a proof, which is based on a topological argument.
In this paper, we give a different, more self-contained proof.
The fact that the proof uses the recursion theorem should also be of some interest.\end{remark}

\begin{proof}  
($\Longleftarrow$).  We give an algorithm that can decide for each 
coalition whether it is winning or not: Given is a characteristic 
index~$e$ of a coalition~$S$.  Generate the elements of $T_0$ and 
$T_1$; we can do that effectively since these sets are r.e. Wait 
until an initial segment of~$S$ is generated.  (Since a characteristic 
index is given, we can decide whether a string generated is an initial 
segment of~$S$.)  If the initial segment is in~$T_0$, then $S$~is 
losing; if it is in~$T_1$, then $S$~is winning.

\bigskip

($\Longrightarrow$).  Suppose $\omega$~is $\delta$-computable.  Let 
$\delta'$ be a p.r.\ extension of $\delta_{\omega}$; such a $\delta'$ 
exists since $\omega$~is $\delta$-computable.

\emph{Overview.} 

From Proposition~\ref{cutprop}, our goal is to  \emph{effectively enumerate}  
a determining initial segment~$S\cap k$ of each losing coalition~$S$ in~$T_0$ 
and that of each winning coalition in $T_1$.

We will define a certain recursive function~$y(e)$ in Step~1.  
In Step~2, we will first define the sets $T_i$, where $i\in\{0,1\}$, as the 
collection of certain strings (of 0's and 1's) of length~$k(e)$ (to be 
defined) for those $e\in\N$ satisfying $\delta'(y(e))=i$.  
In particular, $T_0\cup T_1$~includes, for each 
characteristic index~$e$, the $k(e)$-initial segment of the recursive 
coalition indexed by~$e$.  We will then show that $T_0$ 
and~$T_1$ satisfy the conditions stated.

We use the following \textbf{notation}.  We write $\varphi_{e,s}(x)=y$ if $x$, 
$y$, $e<s$ and $y$ is the output of $\varphi_e(x)$ in less than $s$ 
steps of the $e$th Turing program~\citep[p.~16]{soare87}.  
In particular, if $e$th Turing program does not give an output for $x$ in $\leq s$ steps,  
then $\varphi_{e,s}(x)$ is undefined.
We fix a Turing program for~$\delta'$ and denote by $\delta'_s(y)$ the 
computation of $\delta'(y)$ up to step~$s$ of the program.

\bigskip 

\emph{Step~1.  Defining a recursive function~$y(e)$.}

We define a recursive function~$f(e,y)$ in Step~1.1.  
In Step~1.2, we apply a variant of the Recursion Theorem to 
$f(e,y)$ and obtain~$y(e)$.

\medskip

\emph{Step~1.1.  Defining a recursive function $f(e,y)$.}

Define an r.e.\ set $Q_0\subseteq\N$ by
$y\in Q_0$ iff there exists~$s$ such that
$\delta'_{s}(y)=0$ or $\delta'_{s}(y)=1$.
Define a p.r.\ function 
\[
s_0(y)=\mu s [\delta'_{s}(y)\in\{0,1\}],
\]
which converges for $y\in Q_0$.  

Fix a recursive set~$\F$ of characteristic indices for finite sets 
such that each finite set has at least one characteristic index 
in~$\F$.  (An example of $\F$ is the set consisting of the code 
numbers (G\"odel numbers) of the Turing programs of a particular 
form.)  For $s\in\N$, let $\F_s=\F\cap s$ be the finite set of numbers 
$e<s$ in~$\F$.

Define a set $Q_1\subseteq\N\times\N$ by $(e,y)\in Q_1$ iff (i)~$y\in 
Q_0$ and (ii)~there exist $s'\geq s_0:=s_0(y)$ and $e'\in\F_{s'}$ such 
that (ii.a)~$\delta'_{s'}(e')=1-\delta'_{s_0}(y)$ and that 
(ii.b)~$\varphi_{e',s'}$ is an extension of $\varphi_{e,s_0-1}$.\footnote{\label{note:extension}
Condition~(ii.b), written $\varphi_{e',s'}\supseteq\varphi_{e,s_0-1}$,
means that if $\varphi_{e,s_0-1}(z)=u$, then $\varphi_{e',s'}(z)=u$.
In particular, if $\varphi_{e,s_0-1}(z)$ is undefined, then $\varphi_{e',s'}(z)$ can take any value or no value.
It is possible that $\varphi_{e,s_0-1}$ is undefined (has a ``hole'') for $z$ but defined for some $z'>z$.} %
Note that if $(e,y)\in Q_1$, then 
$s_0=s_0(y)$ is defined and $\delta'_{s_0}(y)\in \{0,1\}$.  We can 
easily check that $Q_1$ is r.e. Given $(e,y)\in Q_1$, let $s_1$ be the 
least $s'\geq s_0$ such that conditions (ii.a) and (ii.b) hold for 
some $e'\in\F_{s'}$.  Let $e_0$ be the least $e'\in\F_{s'}$ such that 
conditions (ii.a) and (ii.b) hold for $s'=s_1$.  We can view $e_0$ as 
a p.r.\ function $e_0(e,y)$, which converges for $(e,y)\in Q_1$.

Define a partial function~$\psi$ by 
\[
\psi(e,y,z) =\left\{ 
	\begin{array}{ll}
		\varphi_{e_0}(z) & \mbox{if $y\in Q_0$ and $(e,y)\in Q_1$},\\
		\varphi_{e,s_0-1}(z) & \mbox{if $y\in Q_0$ and $(e,y)\notin Q_1$},\\
		\varphi_e(z) & \mbox{if $y\notin Q_0$}.
	\end{array}
	\right.
\]

\begin{lemma}
$\psi$ is p.r.
\end{lemma}

\begin{subproof}
We show there is a sequence of p.r.\ functions~$\psi^s$
such that $\psi=\bigcup_s \psi^s$.  We then apply the Graph Theorem to conclude $\psi$ is p.r.

\medskip

For each $s\in \N$, define a recursive set $Q_0^s\subseteq\N$ by
$y\in Q_0^s$ iff there exists~$s'\leq s$ such that
$\delta'_{s'}(y)=0$ or $\delta'_{s'}(y)=1$.  We have $y\in Q_0$ iff $y\in Q_0^s$ for some~$s$.
Note that if $y\in Q_0^s$, then $s\ge s_0=s_0(y)$.

For each $s\in \N$, define a recursive set $Q_1^s\subseteq\N\times\N$ by 
$(e,y)\in Q_1^s$ iff (i)~$y\in Q_0^s$ and 
(ii)~there exist $s'$ such that $s_0:=s_0(y)\le s' \le s$ and $e'\in\F_{s'}$ such 
that (ii.a)~$\delta'_{s'}(e')=1-\delta'_{s_0}(y)$ and that 
(ii.b)~$\varphi_{e',s'}$ is an extension of $\varphi_{e,s_0-1}$.  
(Conditions (ii.a) and~(ii.b) are the same as those in the definition of~$Q_1$.)
We have $(e,y)\in Q_1$ iff $(e,y)\in Q_1^s$ for some $s$.
Note that if $(e,y)\in Q_1^s$, then $s_0\le s_1 \le s$.

For each $s\in \N$, define a p.r.\ function~$\psi^s$ by 
\[
\psi^s(e,y,z) =\left\{ 
	\begin{array}{ll}
		\varphi_{e_0,s}(z) & \mbox{if $y\in Q_0^s$ and $(e,y)\in Q_1^s$},\\
		\varphi_{e,s_0-1}(z) & \mbox{if $y\in Q_0^s$ and $(e,y)\notin Q_1^s$},\\
		\varphi_{e,s}(z) & \mbox{if $y\notin Q_0^s$}.
	\end{array}
	\right.
\]

\emph{We claim that $\bigcup_s \psi^s$ is a partial function 
(i.e., $\bigcup_s \psi^s(e,y,z)$ does not take more than one value)
and that $\psi=\bigcup_s \psi^s$}:

\begin{itemize}
\item Suppose $y\notin Q_0$.  Then $y\notin Q_0^s$ for any~$s$.  So, for all $s$, $\psi^s(e,y,z)=\varphi_{e,s}(z)$.
Hence $\bigcup_s \psi^s(e,y,z)=\varphi_{e}(z)=\psi(e,y,z)$ as desired.

\item Suppose ($y\in Q_0$ and) $(e,y)\in Q_1$.  Then $s_0$, $s_1$, and $e_0$ are defined and $s_1\ge s_0$.
If $s<s_0$, then since $y\notin Q_0^s$, we have $\psi^s(e,y,z)=\varphi_{e,s}(z)$.
If $s_0\le s<s_1$, then since $y\in Q_0^s$ and $(e,y)\notin Q_1^s$, we have 
$\psi^s(e,y,z)=\varphi_{e,s_0-1}(z)$.
If $s_1\le s$, then since $y\in Q_0^s$ and $(e,y)\in Q_1^s$, we have 
$\psi^s(e,y,z)=\varphi_{e_0,s}(z)$.
Hence $\bigcup_s \psi^s(e,y,z)=\varphi_{e,s_0-1}(z)\cup (\bigcup_{s\ge s_1}\varphi_{e_0,s}(z))$.
The definition of $s_1$ implies that when $s_1\le s$, 
$\varphi_{e,s_0-1} \subseteq \varphi_{e_0,s_1}\subseteq \varphi_{e_0,s}$.
Thus $\bigcup_s \psi^s(e,y,z)=\bigcup_{s\ge s_1}\varphi_{e_0,s}(z)=
\varphi_{e_0}(z)=\psi(e,y,z)$ as desired.

\item Suppose $y\in Q_0$ and $(e,y)\notin Q_1$.  Then $s_0$ is defined.
If $s<s_0$, then since $y\notin Q_0^s$, we have $\psi^s(e,y,z)=\varphi_{e,s}(z)$.
If $s_0\le s$, then since $y\in Q_0^s$ and $(e,y)\notin Q_1^s$, we have 
$\psi^s(e,y,z)=\varphi_{e,s_0-1}(z)$.
Hence $\bigcup_s \psi^s(e,y,z)=\varphi_{e,s_0-1}(z)=\psi(e,y,z)$ as desired.
\end{itemize}

Define a partial function~$\hat{\psi}$ by $\hat{\psi}(s,e,y,z)=\psi^s(e,y,z)$.
Then from the construction of $\psi^s$, $\hat{\psi}$ is p.r.\ by Church's Thesis.
By the Graph Theorem, the graph of~$\hat{\psi}$ is r.e.

\emph{We claim that $\psi=\bigcup_s \psi^s$ is p.r.}  
By the Graph Theorem it suffices to show that its graph is r.e.  We have 
$(e,y,z,u)\in \psi  \Longleftrightarrow  \exists s\ (e,y,z,u)\in \psi^s 
 \Longleftrightarrow \exists s\ (s,e,y,z,u)\in \hat{\psi}$.
Since the graph of~$\hat{\psi}$ is r.e., it follows that the graph of~$\psi$ is r.e.\end{subproof}

\bigskip

Since $\psi$ is p.r., there is a recursive function~$f(e,y)$ such that
$\varphi_{f(e,y)}(z)=\psi(e,y,z)$ by the Parameter Theorem.

\medskip

\emph{Step~1.2.  Applying the Recursion Theorem to obtain $y(e)$.}

Since $f(e,y)$ is recursive, by the Recursion Theorem with Parameters \citep[p.~37]{soare87} 
there is a recursive function~$y(e)$ such that $\varphi_{y(e)}=\varphi_{f(e,y(e))}$.  So, we have
$\varphi_{y(e)}(z)=\psi(e,y(e),z)$.

\emph{We claim that $y=y(e)$ cannot meet the first case ($y\in Q_0$ and $(e,y)\in Q_1$)
in the definition of $\psi$}.  Suppose $y(e)\in Q_0$ and $(e,y(e))\in Q_1$.
Since $\varphi_{y(e)}(z)=\psi(e,y(e),z)$, by the definition of~$\psi$ we have
on the one hand $\varphi_{y(e)}=\varphi_{e_0}$.
By~(ii.a) of the definition of~$Q_1$ and by the definition of~$e_0$, we have
on the other hand $\delta'(e_0)=1-\delta'(y(e)) \neq \delta'(y(e))$.
This contradicts the fact that $\delta'$ extends the $\delta$-indicator~$\delta_\omega$.

Therefore, we can express $\varphi_{y(e)}(z)=\psi(e,y(e),z)$ as follows:\footnote{
Roughly speaking, the first case in the definition of~$\psi$ corresponds to the existence of an
index $e'$ for a finite set extending $\varphi_{e,s_0-1}$ such that 
$\delta'(e')$ is different from $\delta'(y(e))$.
Observe that $y(e)$ is defined so that 
this case (as just shown) as well as  the third case (Step 2.3 below) in the definition of~$\psi$
will not occur in (\ref{varphi-y}) if $e$ is a characteristic index.
It follows that if $e$ is a characteristic index, then 
all finite coalitions extending $\varphi_{e,s_0-1}$ have the same winning/losing status
(the $\delta'$-value of their indices is the same as $\delta'(y(e))$).
Thus $\varphi_{e,s_0-1}$ is like a determining string---except that it may fail to be a string because of ``holes'' as discussed in footnote~\ref{note:extension}.} 
\begin{equation} \label{varphi-y}
\varphi_{y(e)}(z) =\left\{ 
	\begin{array}{ll}
		\varphi_{e,s_0-1}(z) & \mbox{if $y(e)\in Q_0$ ($(e,y(e))\notin Q_1$ implied)},\\
		\varphi_e(z) & \mbox{if $y(e)\notin Q_0$}.
	\end{array}
	\right.
\end{equation}

\bigskip

\emph{Step~2  Defining $T_0$ and~$T_1$ and verifying the conditions.}

For $i\in \{0,1\}$, let $T_i$ be the collection of all the strings~$\tau$ of length~$k(e):=s_0-1$
(where $s_0=s_0(y(e))$) that extends $\varphi_{e,s_0-1}$ for all those~$e$ such that $\delta'(y(e))=i$.
(Note that $\delta'(y(e))\in \{0,1\}$ iff $y(e)\in Q_0$.) 
We show $T_0$ and $T_1$ satisfy the conditions.

\medskip

\emph{Step~2.1.  $T_0$ and $T_1$ are r.e.}

This is obvious since $s_0$, $\delta'$, and $y$ are p.r.
(In other words, for each $e$,  first find whether $\delta'(y(e))\in \{0,1\}$. If not, we do not enumerate
any segment in $T_0$ or $T_1$.  If  $\delta'(y(e))=i \in \{0,1\}$, then we
have  corresponding strings whose length is effectively obtained. So we enumerate
them in $T_i$.  This procedure ensures that $T_0$ and $T_1$ are r.e.)

\medskip

\emph{Step~2.2.  $T_0$ and $T_1$ consist of determining strings.}

We show that $T_0$ consists of losing determining strings.  
We can show that $T_1$ consists of winning determining strings in a similar way.

Suppose $\delta'(y(e))=0$.  Then $y(e)\in Q_0$.  Let $s_0=s_0(y(e))$ and $k(e)=s_0-1$.
Since $(e,y(e))\notin Q_1$ by (\ref{varphi-y}), 
there is no $e'\in \F$ such that $\delta'(e')=1-\delta'(y(e))=1$ and that 
$\varphi_{e'}$ is an extension of $\varphi_{e,s_0-1}$.  
(Note that $\delta'(e')\in \{0,1\}$ if $e'\in \F$.)
Hence any finite coalition that extends $\varphi_{e,s_0-1}$ is losing.

Therefore,  all  strings~$\tau$ in $T_0$
(i.e., all the finite strings of length~$k(e)$ that extend $\varphi_{e,s_0-1}$ 
for some~$e$ such that $\delta'(y(e))=0$) are losing determining for \emph{finite} coalitions.
By Proposition~\ref{cutprop}~(iii), all strings~$\tau$ in~$T_0$ are losing determining strings.

\medskip

\emph{Step~2.3.  Any coalition has an initial segment in $T_0\cup T_1$.}

Let $S$ be a coalition, which is recursive.  Pick a characteristic index $e$  for~$S$.
We first show that $\delta'(y(e))\in \{0,1\}$ (i.e., $y(e)\in Q_0$).
Suppose $y(e)\notin Q_0$.  By (\ref{varphi-y}),
we have $\varphi_{y(e)}=\varphi_e$.  So $y(e)$~is a characteristic index for~$S$.
Hence  $\delta'(y(e))\in \{0,1\}$.  That is, $y(e)\in Q_0$, which is a contradiction.

By the definitions of $T_0$ and $T_1$, since $\delta'(y(e))\in \{0,1\}$,
all the strings of length $k(e)=s_0-1$ extending $\varphi_{e,s_0-1}$
are in $T_0\cup T_1$.   In particular, since the $k(e)$-initial segment $S\cap k(e)$ of 
the characteristic function $\varphi_e$ for~$S$ extends $\varphi_{e,s_0-1}$,
the initial segment $S\cap k(e)$ is in~$T_0\cup T_1$.\end{proof}

\section{Applications: Finite Carriers, Finite Winning Coalitions, Prefilters, and Nonanonymity}
\label{applications}

Theorem~\ref{delta0det} is a powerful theorem.
We can obtain as its corollaries some of the results in \citet{mihara04mss}.

\subsection{Finite carriers}

The following proposition asserts that games that are essentially finite satisfy 
$\delta$-computability, as might be expected.  
We give here a proof that uses the characterization theorem.

\begin{prop}[Mihara, 2004, Proposition~5] \label{d0pos} 
	Suppose that a simple game~$\omega$ has a finite carrier.
	Then $\omega$ is $\delta$-computable.
\end{prop}

\begin{proof} 
Suppose that $\omega$ has a finite carrier~$S$.
Let $k=\max S+1$ (we let $k=0$ if $S=\emptyset$).
Let \[
T_1=\{\tau: \textrm{$\tau$ is a string of length~$k$ and $\tau\in\omega$} \}
\]
(where $\tau\in\omega$ means that 
the set $\{i<k: \tau(i)=1\}$ represented by $\tau$, viewed as a characteristic function,
is in $\omega$)
and $T_0=\{\tau: \textrm{$\tau$ is a string of length~$k$ and $\tau\notin\omega$} \}$.
We verify the conditions of Theorem~\ref{delta0det}.

Since  $T_0$ and $T_1$ are finite, they are r.e.  
Since $T_0\cup T_1$ consists of all strings of length~$k$, any coalition has an initial segment in it.

We show that $T_1$ consists of winning determining strings.
(We can show that $T_0$ consists of losing determining strings in a similar way.)
Suppose $G\cap k=\tau\in T_1$.  It suffices to show that $G\in \omega$.
By the definition of $T_1$, we have $\tau\in \omega$.  
This implies $\tau\cap S\in \omega$ since $S$ is a carrier.
Since $\tau\cap S=(G\cap k)\cap S=G\cap(k\cap S)=G\cap S$, it follows that
$G\cap S\in \omega$.  Since $S$ is a carrier, we get $G\in\omega$.\end{proof}

In particular, \emph{if a simple game~$\omega$ is dictatorial, then $\omega$ 
is $\delta$-computable}.  Indeed, the coalition 
consisting of the dictator is a finite carrier for the dictatorial 
game~$\omega$.

\subsection{Finite winning coalitions}\label{sec:finwin}

Note in Proposition~\ref{d0pos} that if a game has a finite carrier~$S$ and $N$ is winning,
then there exists a finite winning coalition, namely $S=N\cap S$.  
When there does not exist a finite 
winning coalition, it is a corollary of the following negative result---itself 
a corollary of Theorem~\ref{delta0det}---that the computability condition is violated.

\begin{prop}[Mihara, 2004, Proposition~6]\label{d0neg1} 
	Suppose that a simple game~$\omega$ has an infinite winning 
	coalition~$S\in\omega$ such that for each $k\in\N$, its $k$-initial 
	segment $S\cap k$ is losing.  Then $\omega$ is not $\delta$-computable.
\end{prop}

\begin{proof}
Suppose that  $\omega$ is $\delta$-computable.  Since $S$ is winning, 
by Lemma~\ref{cutlemma} (or by Proposition~\ref{cutprop} or by Theorem~\ref{delta0det})
there is some $k\in \N$ such that $G=S\cap k$ is winning.
This contradicts the assumption of the proposition.\end{proof}

Theorem~\ref{delta0det} immediately gives the following extension of 
Corollary~7 of \citet{mihara04mss}.  It gives a useful criterion for 
checking computability of simple games.  
Here, a \emph{cofinite} set is the complement of a finite set.

\begin{prop}\label{d0neg1cor} 
	Suppose that a $\delta$-computable simple game has a winning coalition.
	Then, it has both finite winning coalitions and cofinite winning coalitions.
\end{prop}

We also prove a result that is close to Proposition~\ref{d0neg1}.

\begin{prop}[Mihara, 2004, Proposition~8] \label{d0neg2} 
Suppose that $\emptyset\notin\omega$.  Suppose that the simple 
game~$\omega$ has an infinite coalition~$S\in\omega$ such that for 
each $k\in\N$, its difference $S\setminus k=\{\,s\in S: s\geq k\,\}$ from the 
initial segment is winning.  Then $\omega$ is not  $\delta$-computable.
\end{prop}

\begin{proof}
Suppose $\omega$ is  $\delta$-computable.
Since $\emptyset\notin\omega$, there is a losing determining string
$\tau=00\cdots0$ of length~$k$ by Theorem~\ref{delta0det}.  
By assumption, $S\setminus k$ is winning.
But $(S\setminus k)\cap k=\tau$ and that $\tau$ is a losing determining string imply that
$S\setminus k$ is losing, which is a contradiction.\end{proof}

Again, the following proposition gives a useful criterion for 
checking computability of simple games.

\begin{prop}\label{d0neg2cor} 
	Suppose that a $\delta$-computable simple game has a losing coalition.
	Then, it has both finite losing coalitions and cofinite losing coalitions.
\end{prop}

\subsection{Prefilters, filters, and ultrafilters}

From the propositions in Section~\ref{sec:finwin}, examples of a noncomputable simple game are easy to come by.

\begin{example}\label{q-compl} 
For any $q$ satisfying $0< q \leq \infty$, let $\omega$ be the \textbf{$q$-complement rule} defined as follows:  
$S\in \omega$ if and only if $\#(N\setminus S)< q$.  For example, if $q=1$, then the $q$-complement rule is the unanimous game, consisting of $N$ alone.  If $q=\infty$, then the game consists of cofinite coalitions 
(the complements of finite coalitions).  Proposition~\ref{d0neg1cor}
implies that \emph{$q$-complement rules are not $\delta$-computable}, since they have no finite winning coalitions.
Any $q$-complement rule is a prefilter and it is a monotonic, proper, nonstrong, and anonymous simple game.  
If $q>1$, it is nonweak, but any finite intersection of winning coalitions is nonempty
 (i.e., has an infinite Nakamura number, to be defined in Section~\ref{core}).
Note that if $1<q<\infty$, then the $q$-complement rule is not a filter since it is not closed with respect to finite intersection.\end{example}

Example~\ref{q-compl} gives examples of a prefilter that is not a filter.  
It also gives two examples of a filter that is not an ultrafilter:
the unanimous game is a principal filter and 
the game consisting of all cofinite coalitions is a nonprincipal filter.
\citet{mihara01scw} gives a constructive example of an ultrafilter.

Some prefilters are computable, but that is true only if they have a veto player:
according to Proposition~\ref{filter-nakamura} below, if a prefilter
is $\delta$-computable, then it is weak.
 
If $\omega$ is a filter, then it is $\delta$-computable
if and only if it is has a finite carrier~\citep[Corollary~11]{mihara04mss}.
In particular, the principal filter $\omega=\{T\in \REC: S\subseteq T\}$
 generated by $S$ has a carrier, namely, $S$.
So, if $S$ is finite, the principal filter is computable.  
If $S$ is infinite, it is noncomputable, since it does not have a finite winning coalition.
For example, the principal filter generated by $S=2\N:=\{0, 2, 4, \ldots \}$ is a monotonic,
proper, nonstrong, weak, and noncomputable simple game.

If $\omega$ is a nonprincipal ultrafilter, it  is not $\delta$-computable by Proposition~\ref{d0neg1cor}, 
since it has no finite winning coalitions (or no cofinite losing coalitions).
It is a monotonic, proper, strong, and nonweak noncomputable simple game.
In fact, an ultrafilter is $\delta$-computable if and only if it is dictatorial~\citep[Lemma~4]{mihara97et}.

\subsection{Nonanonymity}

As an application of the characterization result in Section~\ref{mainresult},  we show that 
$\delta$-computable simple games violate finite anonymity, a weak notion of anonymity.
Proposition~\ref{nonfinanonymous} below strengthens 
an earlier result \citep[Corollary~12]{mihara04mss} about computable games.
The latter result assumes proper, monotonic, $\delta$-computable simple games, 
instead of just $\delta$-computable simple games.

\begin{prop}\label{nonfinanonymous} 
Suppose that $N\in\omega$ and $\emptyset\notin\omega$. 
If the simple game $\omega$ is $\delta$-computable, then it is not finitely anonymous.
\end{prop}

\begin{proof}
Let $\omega$ be a  finitely anonymous $\delta$-computable simple game such that $N\in\omega$ and $\emptyset\notin\omega$. 
Since $N\in\omega$, there is an initial segment $k:=\{0, 1, \ldots, k-1\}=11\cdots1$ (string of 1's of length $k$),
by Lemma~\ref{cutlemma} (or by Proposition~\ref{cutprop} or by Theorem~\ref{delta0det}).
Since $\emptyset\notin\omega$, there is a losing determining string 
$\tau=00\cdots0$ of length~$k'$ by Theorem~\ref{delta0det}.
Then the concatenation $\tau*k=00\cdots011\cdots1$ of $\tau$ and $k$, viewed as a set,
 has the same number of elements as $k$.
Since coalitions $\tau*k$ and $k$ are finite and have the same number of elements, they should be treated equally
by the finitely anonymous $\omega$.  But $\tau*k$  is losing and $k$ is winning.\end{proof}

\section{The Number of Alternatives and the Core}\label{core}

In this section, we apply Theorem~\ref{delta0det} to a social choice problem.
We show (Corollary~\ref{core-nakamura}) 
that computability of a simple game entails a restriction on the number of alternatives
that the set of players (with the coalition structure described by the simple game) can deal with rationally.

For that purpose, we define the notion of a \emph{simple game with (ordinal) preferences},
a combination of a simple game and a set of alternatives and individual preferences.
After defining the \emph{core} for simple games with preferences,
we extend (Theorem~\ref{nakamura-thm}) Nakamura's theorem (\citeyear{nakamura79})
 about the nonemptyness of the core:
the core of a simple game with preferences is always (i.e., for all profiles of preferences) nonempty 
if and only if the number of alternatives is finite and below a certain critical number, 
called the \emph{Nakamura number} of the simple game.
We need to make this extension since what we call a ``simple game'' is not generally 
 what is called a ``simple game'' in \citet{nakamura79}.

We show (Corollary~\ref{nakamura-finite}) that the Nakamura number of a nonweak simple game is finite 
if it is computable, though (Proposition~\ref{nakamura-any}) there is no upper bound for the set
 of the Nakamura numbers of such games.\footnote{
\citet{kumabe-m07nc} study the relations between the Nakamura number and 
computable simple games having various properties.}
It follows from Theorem~\ref{nakamura-thm} that (Corollary~\ref{core-nakamura}) in order for a set of alternatives to always have a maximal element given a nonweak, computable game,
the number of alternatives must be restricted.
In contrast, \emph{some} noncomputable 
(and nonweak) simple games do not have such a restriction (Proposition~\ref{filter-nakamura}),
and in fact have some nice properties.
These results have implications for social choice theory; 
we suggest its connection with the study of Arrow's Theorem (\citeyear{arrow63}).

\subsection{Framework}

Let $N'$ be an arbitrary nonempty set of players and 
$\B\subseteq 2^{N'}$ an arbitrary Boolean algebra of  subsets (called ``coalitions'' in this section) of~$N'$. 
A $\B$-\textbf{simple game} $\omega$ is a subcollection of~$\B$ such that $\emptyset\notin \omega$.
The elements of $\omega$ are said to be winning, and the other elements in~$\B$ 
are losing, as before.
Our ``simple game'' is a $\B$-simple game with $N'=\N$ and $\B=\REC$, if it does not contain~$\emptyset$.
Nakamura's ``simple game'' (\citeyear{nakamura79}) is one with $\B=2^{N'}$.
The properties (such as monotonicity and weakness, defined in Section~\ref{notions}) for simple games are 
redefined for $\B$-simple games in an obvious way.

Let $X$ be a (finite or infinite) set of \textbf{alternatives}, with cardinal number $\#X\geq 2$.
Let $\A$ be the set of \textbf{(strict) preferences}, i.e., 
\emph{acyclic} (for any finite set $\{x_1, x_2, \ldots, x_m\}\subseteq X$,
if $x_1 \succ x_2$, \ldots, $x_{m-1} \succ x_m$, then $x_m \not\succ x_1$;
in particular, $\succ$ is asymmetric and irreflexive) binary relations~$\succ$ on $X$. 
(If $\succ$ is acyclic, we can show that the relation~$\succeq$, defined by 
$x\succeq y \Leftrightarrow y\not\succ x$, is complete, i.e., reflexive and total.)
A \textbf{($\B$-measurable) profile} is a list 
$\p=\profile \in\A^{N'}$ of {individual preferences} $\pref$
such that $\xprefy \in\B$
for all~$x$, $y\in X$.  Denote by $\profs$ the set of all profiles. 

A \textbf{$\B$-simple game with (ordinal) preferences} is a list $(\omega, X, \p)$ of
a $\B$-simple game~$\omega\subseteq\B$, a set $X$ of alternatives, and
a profile $\p=\profile\in \profs$.
Given the $\B$-simple game with preferences, 
we define the dominance relation~$\succ^\p_\omega$ by $x\succ^\p_\omega y$ if and only if there is a winning coalition
$S\in\omega$ such that $x\pref y$ for all $i\in S$.\footnote{
In this definition, $\xprefy$ need not be winning since we do not assume $\omega$ is monotonic.
\citet{andjiga-m00} study Nakamura's theorem,
adopting the notion of dominance that requires the above coalition to be winning.} %
The \textbf{core} $C(\omega, X, \p)$ of the $\B$-simple game with preferences is the set of undominated alternatives:
\[
C(\omega, X, \p)=\{x\in X: \textrm{$\not\!\exists y\in X$ such that $y\succ^\p_\omega x$}\}.
\]

A \textbf{(preference) aggregation rule}  is
a map $\succ \colon \p\mapsto \, \succ^\p$ from profiles~$\p$ of preferences 
to binary relations (social preferences)~$\succ^\p$ 
on the set of alternatives.  For example, the mapping $\succ_\omega$ 
from profiles $\p\in\profs$ of acyclic preferences to dominance relations $\succ^\p_\omega$ 
is an aggregation rule.
We typically restrict individual and social preferences to those binary relations $\succ$ on $X$
that are asymmetric (i.e., complete $\succeq$) and 
either (i)~acyclic or (ii)~transitive (i.e., quasi-transitive $\succeq$) or
(iii)~negatively transitive (i.e., transitive $\succeq$).
An aggregation rule is often referred to as a \emph{social welfare function} when individual preferences
and social preferences are restricted to the asymmetric, negatively transitive relations.

\subsection{Nakamura's theorem and its consequences}


\citet{nakamura79} gives a necessary condition for a $2^{N'}$-simple game with preferences
to have a nonempty core for any profile~$\p$, which is also sufficient if the set $X$ of alternatives is finite.  
To state Nakamura's theorem, we define the \textbf{Nakamura number} $\nu(\omega)$ of a $\B$-simple game~$\omega$ to be the size of the smallest collection of winning coalitions having empty intersection
\[
\nu(\omega)=\min \{\#\omega': \textrm{$\omega' \subseteq \omega $ and $\bigcap\omega'=\emptyset$} \}
\]
if $\bigcap\omega=\emptyset$ (i.e., $\omega$ is nonweak); otherwise, set $\nu(\omega)=\#(2^X)>\#X$.

The following useful lemma \citep[Lemma~2.1]{nakamura79} states that the Nakamura number of
a $\B$-simple game cannot exceed the size of a winning coalition by more than one.

\begin{lemma}\label{nakamura-ceiling} 
Let $\omega$ be a nonweak $\B$-simple game.  
Then $\nu(\omega)\leq \min \{\#S: S\in\omega\}+1$.
\end{lemma}

\begin{proof}
Choose a coalition $S\in \omega$ such that $\#S=\min \{\#S: S\in\omega\}$.
Since $\bigcap\omega=\emptyset$, for each $i\in S$, there is some $S^i\in \omega$ with $i\notin S^i$.
So, $S\cap (\bigcap_{i\in S} S^i)=\emptyset$.  Therefore, $\nu(\omega)\leq \#S+1$.\end{proof}

It is easy to prove \citep[Corollary~2.2]{nakamura79} that the Nakamura number of a nonweak $\B$-simple game is at 
most equal to the cardinal number~$\#N$ of the set of players and that this maximum is attainable if $\B$ 
contains all finite coalitions.
In fact, one can easily construct a computable, nonweak simple game with any given Nakamura number:

\begin{prop}\label{nakamura-any} 
For any integer $k\geq 2$,  there exists a $\delta$-computable, nonweak simple game~$\omega$
with Nakamura number $\nu(\omega)=k$.
\end{prop}

\begin{proof}
Given an integer $k\geq 2$, 
let $S=\{0,1, \ldots, k-1\}$ be a carrier and define $T\in \omega$ iff $\#(S\cap T) \geq k-1$.  
Then $\nu(\omega)=k$.\end{proof}

Since computable, nonweak simple games have winning coalitions, 
it has \emph{finite} winning coalitions by Proposition~\ref{d0neg1cor}.
An immediate corollary of Lemma~\ref{nakamura-ceiling} is the following:
\begin{cor}\label{nakamura-finite} 
Let $\omega$ be a $\delta$-computable, nonweak simple game.  
Then its Nakamura number~$\nu(\omega)$ is finite.
\end{cor}

\citet{nakamura79} proves the following theorem for $\B=2^{N'}$:

\begin{theorem} \label{nakamura-thm} 
Let $\B$ be a Boolean algebra of sets of $N'$.
Suppose that $\emptyset \notin \omega$ and $\omega\neq \emptyset$.
Then the core $C(\omega, X, \p)$ of a $\B$-simple game $(\omega, X, \p)$ with preferences is nonempty for all (measurable) profiles $\p\in \profs$
if and only if $X$ is finite and $\#X<\nu(\omega)$.\end{theorem}

\begin{remark}
At first glance, Nakamura's proof \citep[Theorem~2.3]{nakamura79} of the necessary condition~$\#X<\nu(\omega)$,
does not appear to generalize to an arbitrary Boolean algebra~$\B$:
he constructs certain coalitions from winning coalitions by taking possibly
\emph{infinite unions and intersections}, as well as complements;
a difficulty is that the resulting set of players may not belong to the Boolean algebra~$\B$.
However, it turns out that once we make use of the other necessary condition 
(disregarded by Nakamura) that $X$~is finite, 
we only need to consider \emph{finite} unions and intersections, and his proof actually works.
Since accessible proofs are readily available in the literature \citep[e.g.,][Theorem~3.2]{austensmith-b99}
for $\B=2^{N'}$ and finite sets~$N'$ of players, we choose to relegate the proof to
the working paper \citep[Appendix~A.3]{kumabe-m07csgcc}.  
Unlike others', our proof treats the \emph{measurability} condition ($\p\in \profs$) particularly carefully.\end{remark}

It follows from Theorem~\ref{nakamura-thm} that if a $\B$-simple game $\omega$ is \emph{weak} 
(and satisfies $\emptyset \notin \omega$ and $\omega\neq \emptyset$), 
then the core $C(\omega, X, \p)$ is nonempty for all profiles $\p\in \profs$ 
if and only if $X$~is finite.
The more interesting case is where $\omega$ is nonweak.
Combined with Corollary~\ref{nakamura-finite}, Theorem~\ref{nakamura-thm} has a consequence
for nonweak, computable simple games:

\begin{cor}\label{core-nakamura} 
Let $\omega$ be a $\delta$-computable, nonweak simple game satisfying
$\emptyset \notin \omega$.
Then there exists a finite number $\nu$ (the Nakamura number~$\nu(\omega)$) such that
the core $C(\omega, X, \p)$ is nonempty for all profiles $\p\in \A^{N}_\REC$
if and only if $\#X<\nu$.\end{cor}

If we drop the computability condition, the above conclusion no longer holds.
An example of $\omega$ that has no such restriction on the size of the set~$X$ of alternatives
is a nonweak prefilter (e.g., the $q$-complement rule of Example~\ref{q-compl}, for $q>1$),
which has an infinite Nakamura number.

In fact, we can say more, if we shift our attention from the core---the set of undominated 
alternatives with respect to the dominance relation~$\succ^\p_\omega$---to the 
dominance relation itself. 
(The proof of the following proposition is in the working paper \citep[Appendix~A.4]{kumabe-m07csgcc}.)

\begin{prop}\label{filter-nakamura} 
Let $\omega$ be a nonweak simple game
satisfying $\emptyset \notin \omega$.
\textup{(i)}~$\omega$ cannot be a $\delta$-computable prefilter.
\textup{(ii)}~If $\omega$ is $\delta$-computable; then $\nu(\omega)$ is finite, and
$\succ^\p_\omega$  is acyclic for all $\p\in\A^N_\REC$ if and only if $\#X<\nu(\omega)$.
\textup{(iii)}~If $\omega$ is a prefilter,
then $\succ^\p_\omega$  is acyclic for all $\p\in\A^N_\REC$, regardless 
of the cardinal number $\#X$ of~$X$.\end{prop}

We can strengthen the acyclicity of the dominance relation~$\succ^\p_\omega$ in statement~(iii)
of Proposition~\ref{filter-nakamura} by replacing the statement with one of the following:
(iv)~if $\omega$ is a \emph{filter},
then $\succ^\p_\omega$  is transitive for all $\p$ such that all individuals have transitive preferences~$\pref$;
(v)~if $\omega$ is an \emph{ultrafilter},
then $\succ^\p_\omega$  is asymmetric and negatively transitive for all $\p$ 
such that all individuals have asymmetric, negatively transitive preferences~$\pref$.
In fact, statements (iii), (iv), and (v) each gives an aggregation rule $\succ_\omega\colon \p\mapsto \, \succ^\p$ that satisfies Arrow's conditions of ``Unanimity'' and ``Independence of irrelevant alternatives.''
These results are immediate from the relevant definitions 
(\citet[Proposition~3.2]{armstrong80} gives a proof).
According to Arrow's Theorem (\citeyear{arrow63}), however, if the set $N$ of players were 
replaced by a \emph{finite} set, then social welfare functions given by statement~(v) 
would be dictatorial (and $\omega$ would be weak).

In an attempt to escape from Arrow's impossibility, many authors have investigated the consequences of
relaxing the rationality requirement (negative transitivity of $\succ^\p_\omega$)
for social preferences.
In view of the close connection \citep[Theorems 2.6 and 2.7]{austensmith-b99} 
between the rationality properties of an aggregation rule and preflters 
\citep[also][]{kirman-s72,armstrong80,armstrong85},
Proposition~\ref{filter-nakamura} has a significant implication for this investigation.

\section{Examples}\label{examples}

Propositions~\ref{d0pos}, \ref{d0neg1cor}, and \ref{d0neg2cor} show that the class of computable games
(i)~includes the class of games that have finite carriers and 
(ii)~is included in the class of games that have both finite winning coalitions and cofinite losing coalitions.
In this section, we construct examples showing that these inclusions are strict.

We can find such examples without sacrificing the voting-theoretically desirable properties of simple games.
We pursue this task thoroughly in a companion paper~\citep{kumabe-m06csg64}.
The \emph{noncomputable} simple game example in Section \ref{ex:noncomp} 
that has both finite winning coalitions and cofinite losing coalitions is a sample of that work.
It is monotonic, proper, strong, and nonweak.
Examples of a \emph{computable} simple game that is monotonic, proper, strong, nonweak, 
and has no finite carrier is given in \citet{kumabe-m06csg64,kumabe-m07nc}.

\subsection{A noncomputable game with finite winning coalitions} 
\label{ex:noncomp}

We exhibit here a noncomputable simple game
that is monotonic, proper, strong, nonweak, 
and have both finite winning coalitions and cofinite losing coalitions.
It shows in particular that the class of computable games is strictly smaller than the class 
 of games that have both finite winning coalitions and cofinite losing coalitions.
In this respect, the game is unlike nonweak prefilters
(such as the $q$-complement rules in Example~\ref{q-compl}); 
those examples do not have any finite winning coalitions.
Furthermore, unlike nonprincipal ultrafilters---which are also 
monotonic, proper, strong, and nonweak noncomputable simple games---the 
game is nonweak in a stronger sense: it violates the finite intersection property.

Let $A=N\setminus \{0\}=\{1,2,3,\ldots\}$.  We define the simple game~$\omega$ as follows:
Any coalition except $A^c=\{0\}$ extending the string 1 of length~1 (i.e., any coalition containing~0)  is winning; 
any coalition except $A$ extending the string 0  is losing; $A$ is winning and $A^c$ is losing.
In other words, for all $S\in \REC$,
\[ 
S\in \omega \iff [\textrm{$S=A$ or ($0\in S$ \& $S\neq A^c$)}].
\]

\begin{remark}
The reader familiar with the notion of repeated games (or binary rooted trees) may find the following visualization helpful.
Think of the extensive form of an infinitely repeated game played by you, 
with the stage game consisting of two moves 0 and~1.  
If you choose~1 in the first stage, you will win unless you keep choosing 0 indefinitely thereafter; 
if you choose~0 in the first stage, you will lose unless you keep choosing 1 indefinitely thereafter.
Now, you ``represent'' a certain coalition and play 1 in stage $i$ if $i$ is in the coalition; 
you play~0 in that stage otherwise.
Then the coalition that you represent is winning if you win; it is losing if you lose.
\end{remark}

\begin{lemma}
$\omega$ is not $\delta$-computable.
\end{lemma}

The following proof demonstrates the power of Theorem~\ref{delta0det},
although its full force is not used  (Proposition~\ref{cutprop} suffices).
Proposition~\ref{d0neg1}, which appeared earlier in \citet{mihara04mss}, does not have this power.

\bigskip

\begin{proof}
If $\omega$ is $\delta$-computable, then by Theorem~\ref{delta0det} (or by Proposition~\ref{cutprop}),
$A$ has an initial segment $A\cap k$ that is a winning determining string.
But $A\cap k$ itself is not winning (though it extends the string trivially).
\end{proof}

\begin{lemma}
$\omega$ has both finite winning coalitions and cofinite losing coalitions.
\end{lemma}

\begin{proof}
For instance, $\{0,1\}$ is finite and winning.
$N\setminus \{0,1\}=\{2, 3, 4, \ldots\}$ is cofinite and losing.
\end{proof}

\begin{lemma}
$\omega$ is monotonic.
\end{lemma}

\begin{proof}
Suppose $S\in \omega$ and $S\subsetneq T$.  There are two possibilities.
If $S=A$, then $T=N$, and we have $N\in \omega$ by the definition of $\omega$.
Otherwise, $S$ contains 0 and some other number $i$.  The same is true of $T$, 
implying that $T\in \omega$.
\end{proof}

\begin{lemma}
$\omega$ is proper and strong.
\end{lemma}

\begin{proof}
It suffices to show that $S^c\in\omega \iff S\notin\omega$.
From the definition of $\omega$, we have
\begin{eqnarray*}
S \notin\omega & \iff & \textrm{$S\neq A$ \& ($0\notin S$ or $S=A^c$)} \\
						& \iff & \textrm{$S^c \neq A^c$ \& ($0\in S^c$ or $S^c= A$)} \\
						& \iff & \textrm{($0\in S^c$  \& $S^c \neq A^c$) or $S^c= A$} \\
						& \iff & S^c\in\omega.
\end{eqnarray*}\end{proof}

\begin{lemma}
$\omega$ is not a prefilter.  In particular,  it is not weak.  
\end{lemma}

\begin{proof}
We show that the intersection of some finite family of winning coalitions is empty.
The coalitions $\{0,1\}$, $\{0,2\}$, and~$A$ form such a family.
(Incidentally, this shows that the Nakamura number of $\omega$ is three, since $\omega$ is proper.)\end{proof}

\subsection{A computable game without a finite carrier}\label{ex:nocarrier} 


We exhibit here a computable simple game that does not have a finite carrier. 
It shows that the class of computable games is strictly larger than 
the class of games that have finite carriers.

Our approach is to construct r.e.\ (in fact, recursive) sets $T_0$ and $T_1$ 
of determining strings (of 0's and 1's) satisfying the conditions of Theorem~\ref{delta0det} 
(the full force of the theorem is not needed; the easier direction suffices).
We first give a condition that any string in $T_0\cup T_1$ must satisfy.  
We then specify each of $T_0$ and $T_1$, and construct the simple game by means of these sets.
We conclude that the game is computable by checking (Lemmas~\ref{ex:nocarrier-rec}, \ref{ex:nocarrier-string1},
and \ref{ex:nocarrier-det}) 
that $T_0$ and $T_1$ satisfy the conditions of the theorem.
Finally, we show (Lemma~\ref{ex:nocarrier-none}) that the game does not have a finite carrier.

\bigskip

Let $\{k_s\}_{s=0}^\infty$ be an effective listing (recursive enumeration) of the members of 
the r.e.\ set $\{k : \varphi_k(k)\in \{0,1\}\}$, 
where $\varphi_k(\cdot)$ is the $k$th p.r.\ function of one variable.
We can assume that all elements $k_s$ are distinct.
(Such a listing $\{k_s\}_{s=0}^\infty$ exists by the Listing Theorem~\citep[Theorem~II.1.8 and Exercise II.1.20]{soare87}.)
Thus, 
\[ \CRec \subset \{k : \varphi_k(k)\in \{0,1\}\} = \{k_0, k_1, k_2, \ldots\}, \]
where $\CRec$ is the set of characteristic indices for recursive sets.

Let $l_{0}=k_0+1$, and for $s>0$, let $l_{s}=\max \{l_{s-1}, k_{s}+1\}$.
We have $l_s\geq l_{s-1}$ (that is, $\{l_s\}$ is an nondecreasing sequence of numbers) and 
$l_s>k_s$ for each $s$.  Note also that $l_s\geq l_{s-1}>k_{s-1}$, and $l_s\geq l_{s-2}>k_{s-2}$, etc.\ 
imply that $l_s> k_s$, $k_{s-1}$, $k_{s-2}$, \ldots.

For each $s$, let $F_s$ be the set of strings $\alpha=\alpha(0)\alpha(1)\cdots\alpha(l_s-1)$ 
(the *'s denoting the concatenation are omitted)
of length $l_s$ such that  
\begin{equation} \label{ex:nocarrier1}
\textrm{$\alpha(k_{s})=\varphi_{k_{s}}(k_{s})$ and for each $s'<s$,  $\alpha(k_{s'})=1-\varphi_{k_{s'}}(k_{s'})$.}
\end{equation}
Note that (\ref{ex:nocarrier1}) imposes no constraints on  $\alpha(k_{s'})$ for $s'>s$ and 
no constraints on $\alpha(k)$ for $k\notin\{k_0,k_1,k_2, \ldots\}$, 
while it imposes real constraints for $s'\leq s$,
since $|\alpha|=l_s> k_{s'}$ for such $s'$.
We observe that if $\alpha\in F_s\cap F_{s'}$, then $s=s'$.
 
Let $F=\bigcup_{s}F_s$.  ($F$ will be the union of $T_0$ and $T_1$ defined below.)
We claim that for any two distinct elements $\alpha$ and $\beta$ in $F$
we have neither $\alpha\subseteq \beta$ ($\alpha$ is an initial segment of~$\beta$)
nor $\beta\subseteq\alpha$ 
(i.e., there is $k< \min\{|\alpha|,|\beta|\}$ such that $\alpha(k)\neq \beta(k)$).
 To see this, let $|\alpha|\leq |\beta|$, without loss of generality.  
 If $\alpha$ and $\beta$ have the same length, then the 
 conclusion follows since otherwise they become identical strings.
 If $l_s=|\alpha|< |\beta|=l_{s'}$, then $s<s'$ and by (\ref{ex:nocarrier1}),
 $\alpha(k_{s})=\varphi_{k_{s}}(k_{s})$ on the one hand, but 
$\beta(k_{s})=1-\varphi_{k_{s}}(k_{s})$ on the other hand.  So $\alpha(k_{s})\neq \beta(k_{s})$.

The game $\omega$ will be constructed from the sets $T_0$ and $T_1$ of strings defined as follows:
\begin{eqnarray*}
\alpha\in T_0 & \iff & \exists s \, \textrm{[$\alpha\in F_s$ and $\alpha(k_{s})=\varphi_{k_{s}}(k_{s})=0$]} \\
\alpha\in T_1 & \iff & \exists s \, \textrm{[$\alpha\in F_s$ and $\alpha(k_{s})=\varphi_{k_{s}}(k_{s})=1$].} 
\end{eqnarray*}
We observe that $T_0\cup T_1=F$ and $T_0\cap T_1=\emptyset$.

\emph{Define $\omega$ by $S\in \omega$ if and only if $S$ has an initial segment in $T_1$}.
Lemmas~~\ref{ex:nocarrier-rec}, \ref{ex:nocarrier-string1}
and \ref{ex:nocarrier-det} establish computability of $\omega$
by way of Theorem~\ref{delta0det}.

\begin{lemma} \label{ex:nocarrier-rec}
$T_0$ and $T_1$ are recursive.
\end{lemma}

\begin{proof}
We prove that $T_0$ is recursive; the proof for $T_1$ is similar.
We give an algorithm that can decide for each given string whether it is in $T_0$ or not.

To decide whether a string $\sigma$ is in $T_0$, generate $k_0$, $k_1$, $k_2$, \ldots,
compute $l_0$, $l_1$, $l_2$, \ldots, and determine $F_0$, $F_1$, $F_2$, \ldots
until we find the least $s$ such that $l_s\geq |\sigma|$.
If $l_s > |\sigma|$, then $\sigma\notin F_s$.
Since $l_s$ is nondecreasing in $s$ and $F_s$ consists of strings of length~$l_s$, 
it follows that $\sigma\notin F$, implying $\sigma\notin T_0$.

If $l_s= |\sigma|$, then check whether $\sigma \in F_s$; this can be 
done since the values of $\varphi_{k_{s'}}(k_{s'})$ for $s'\leq s$ in (\ref{ex:nocarrier1})
are available and $F_s$ determined by time $s$.
If $\sigma\notin F_s$ and $l_{s+1}>l_s$, then $\sigma \notin T_0$ as before.
Otherwise check whether $\sigma \in F_{s+1}$.  
If $\sigma \notin F_{s+1}$ and $l_{s+2}>l_{s+1}=l_s$, then $\sigma \notin T_0$ as before.
Repeating this process, we either get $\sigma\in F_{s'}$ for some $s'$ or 
$\sigma\notin F_{s'}$ for all $s'\in \{s': l_{s'}=l_s\}$.
In the latter case, we have $\sigma \notin T_0$.
In the former case, if $\sigma(k_{s'})=\varphi_{k_{s'}}(k_{s'})=1$, 
then $\sigma \in T_1$ by the definition of $T_1$; hence it is not in $T_0$.
Otherwise $\sigma(k_{s'})=\varphi_{k_{s'}}(k_{s'})=0$, 
and we have $\sigma \in T_0$.\end{proof}

\begin{lemma}\label{ex:nocarrier-string1}
$T_1$ consists only of winning determining strings for $\omega$;
$T_0$ consists only of losing determining strings for $\omega$.
\end{lemma}

\begin{proof}
Let $\alpha\in T_1$.  If a coalition~$S$ extends $\alpha$, then by the definition of~$\omega$,
$S$~is winning.  This proves that $\alpha$~is a winning determining string.

Let $\alpha\in T_0$.  Suppose a coalition~$S$ extends $\alpha\in T_0\subset T_0\cup T_1=F$.
If $\beta\in F$ and $\beta\neq \alpha$, 
we have, as shown before, $\alpha \not \subseteq \beta$ and $\beta\not \subseteq\alpha$,
which implies that $S$~does not extend~$\beta$.  So, in particular, $S$ does not extend any
string in~$T_1$.  It follows from the definition of~$\omega$ that $S$~is losing.
This proves that $\alpha$~is a losing determining string.\end{proof}

\begin{lemma} \label{ex:nocarrier-string}
For each $s$, any string $\alpha$ of length $l_s$ such that $\alpha(k_s)=\varphi_{k_s}(k_s)$ 
extends a string in $\bigcup_{t\leq s}F_t$.
\end{lemma}

\begin{proof}
We proceed by induction on $s$.
Let $\alpha$ be a string of length $l_s$ such that $\alpha(k_s)=\varphi_{k_s}(k_s)$. 
If $s=0$, we have $\alpha\in F_0$; hence the lemma holds for $s=0$. 
Suppose the lemma holds for $s'<s$. If for some $s'<s$, $\alpha(k_{s'})=\varphi_{k_{s'}}(k_{s'})$,
then by the induction hypothesis, the $l_{s'}$-initial segment $\alpha \cap l_{s'}$ of $\alpha$
extends a string in $\bigcup_{t\leq s'}F_t$. So $\alpha$  extends a string in $\bigcup_{t\leq s}F_t$.  
Otherwise,  we have for each $s'<s$,  $\alpha(k_{s'})=1-\varphi_{k_{s'}}(k_{s'})$. 
Then by (\ref{ex:nocarrier1}), $\alpha\in F_s\subset \bigcup_{t\leq s}F_t$.\end{proof}

\begin{lemma} \label{ex:nocarrier-det}
Any coalition $S\in\REC$ has an initial segment in $T_0$ or $T_1$.
\end{lemma}

\begin{proof}
Suppose $\varphi_k$ is the characteristic function for a recursive coalition~$S$. 
Then $k\in\{k_0,k_1,k_2, \ldots\}$ since this set contains the set $\CRec$ of characteristic indices.
So $k=k_s$ for some $s$.  Consider the initial segment $S\cap l_s$.  It extends a string in
$\bigcup_{t\leq s}F_t$ by Lemma~\ref{ex:nocarrier-string}.  The conclusion follows 
since $\bigcup_{t\leq s}F_t \subset F=T_0\cup T_1$.\end{proof}

\begin{lemma} \label{ex:nocarrier-none}
$\omega$ does not have a finite carrier.
\end{lemma}

\begin{proof}
We will construct a set~$A$ such that for infinitely many $l$,
the $l$-initial segment $A\cap l$ has an extension (as a string) that is winning 
and for infinitely many $l'$, $A\cap l'$ has an extension that is losing.  
This implies that $A\cap l$ is not a carrier of~$\omega$ for any such~$l$.
So no subset of $A\cap l$ is a carrier.  Since there are arbitrarily large such~$l$, 
this proves that $\omega$ has no finite carrier.

Let $A$ be a set such that for each $k_t$, $A(k_t)=1-\varphi_{k_t}(k_t)$. 
For any $s'>0$ and $i \in \{0,1\}$, there is an $s>s'$ such that $k_s> l_{s'}$ and $\varphi_{k_s}(k_s)=i$.

For a temporarily chosen $s'$, fix $i$ and fix such $s$.  Then choose the greatest $s'$ satisfying these conditions.
Since $l_s>k_s>l_{s'}$, there is a string $\alpha$ of length~$l_s$ extending (as a string) $A\cap l_{s'}$ 
such that $\alpha\in F_s$.  
Since $\alpha(k_s)=\varphi_{k_s}(k_s)=i$, we have $\alpha\in T_i$.  

There are infinitely many such $s$, so there are infinitely many such $s'$. 
It follows that for infinitely many $l_{s'}$, the initial segment $A\cap l_{s'}$ is a substring 
of some  string~$\alpha$ in $T_1$ (by Lemma~\ref{ex:nocarrier-string1}, $\alpha$ is winning in this case), 
and for infinitely many $l_{s'}$, $A\cap l_{s'}$ is a substring of some (losing) string~$\alpha$ in $T_0$.\end{proof}

\section*{Acknowledgements}

We would like to thank an anonymous referee for useful suggestions.  %
This is an outcome of a  long-term collaboration started in 1998.  
It could not have been produced without a lack of grants for shorter-term projects.







\newpage

\appendix

\section{Not in the JME Version} \label{not-in-JME} 

This appendix collects details omitted from the version published in \emph{Journal of Mathematical Economics}.

\subsection{Multiple-choice or essay: A remark on the notion of computability}

One might argue that the scenario preceding the definition of $\delta$-computability in Section~\ref{comp:notions} makes 
the aggregator's task more difficult than need be.  
The difficulty comes from the fact that, like an essay exam, there is too much freedom on the side of the inquirer, the argument would go, in the sense that each recursive coalition has infinitely many indices and that 
the index presented may be an illegitimate one.  
An alternative notion of computability that deals with these problems might use
a ``multiple-choice format,'' in which the aggregator gives possible indices that the inquirer can choose from. %
Unfortunately, such a ``multiple-choice format'' would not work as one might wish.  

Indeed, we claim that 
\emph{there is no effective listing $e_0$, $e_1$, $e_2$, \ldots of characteristic indices such that 
for each recursive coalition $S$ there is at 
least one $e_i$ that represents the coalition (i.e., $e_i$ is a characteristic index for $S$)}.
To prove this claim, suppose there is such a listing and let $S$ be the set defined by 
$i\in S$ if and only if $\varphi_{e_i}(i)=0$.
Then since $\varphi_{e_i}(i)\downarrow$ for any $i$, we have $S$ recursive.  On the other hand, the characteristic 
function for $S$ is not equal to any $\varphi_{e_i}$.  To see this, suppose that it is equal to $\varphi_{e_i}$.
Then, if $i\in S$, we have $\varphi_{e_i}(i)=1$, the definition of~$S$ then implies $i\notin S$, a contradiction; 
if $i\notin S$, we have a similar contradiction.  The claim is thus proved.

Given this impossibility result, one might wish to relax the condition and
allow some $e_i$ in the listing to fail to be a characteristic index.
Adopting a notion of computability based on such a listing is a halfway solution, 
fitting into neither the essay-exam scenario nor the multiple-choice alternative to it.

\subsection{Another proof of Theorem \ref{delta0det}}

We derive Theorem~\ref{delta0det} from a result in \citet{kreisel-ls59}
and \citet{ceitin59}.  In this proof we largely follow the terminology of 
\citet[pages~186--192 and 205--210]{odifreddi92}, who gives a topological argument.
In this proof only, a \emph{string} refers to a finite sequence $\sigma=\sigma(0)\sigma(1)\cdots\sigma(k)$
of natural numbers (not necessarily $0$ or~$1$).

Let $\mathcal{PR}$ be the class of partial recursive (unary) functions and $\mathcal{R}$ the class of recursive functions.  
An \emph{effective operation on $\mathcal{R}$} is a functional (function)~$F\colon \mathcal{R}\to \mathcal{R}$
such that for some partial recursive function~$\psi$,
\[
\varphi_e\in\mathcal{R}\Longrightarrow [\textrm{$\psi(e)\downarrow$ and $F(\varphi_e)=\varphi_{\psi(e)}$}].
\]

We introduce a topology into the set of partial (unary) functions by viewing it as a product space $S^{\N}$, 
with $S=\N\cup\{\uparrow\}$, $\uparrow$ being a distinguished element for the undefined value. 
For a string $\sigma$, let $A_{\sigma}=\{f\in \mathcal{R}: \textrm{if $\sigma(x)\downarrow$, then $f(x)=\sigma(x)$}\}$ be the set of recursive functions that extend~$\sigma$.  These sets~$A_\sigma$ are the basic open sets.
Let $t$ be a recursive bijection between the set $\N$ of natural numbers and the set of strings. 
We say a continuous functional $F\colon \mathcal{PR}\to \mathcal{PR}$ is \emph{effectively continuous on $\mathcal{R}$} if $F$ maps $\mathcal{R}$ to~$\mathcal{R}$ and for some  recursive function~$\psi$,
\begin{equation}\label{effectcont}
F^{-1}(A_{\sigma})=\{f: \textrm{$f\in A_{\nu}$ for some $\nu\in\{t(a) : a\in W_{\psi(t^{-1}(\sigma))}\}$}\}
\end{equation}
(requiring that the open sets~$F^{-1}(A_{\sigma})$ be obtained in a certain effective way).

\citet{kreisel-ls59} and \citet{ceitin59} prove the
theorem~\citep[Theorem II.4.6]{odifreddi92} stating that 
\emph{the effective operations on $\mathcal{R}$ are exactly 
the restrictions of the effectively continuous functionals on~$\mathcal{R}$}. 

In our context, suppose that $\omega$ is a $\delta$-computable simple game.  
Let $\delta'$ be a p.r.\ extension of $\delta_{\omega}$.
Further let $\delta''$ be such that $\delta'(x)\uparrow \, \Leftrightarrow \delta''(x)\uparrow$ and 
$\delta'(x)=i\Leftrightarrow \delta''(x)=e_i$, 
where for each $i\in\N$, $e_i$ is an index of the constant (recursive) function whose value is always~$i$.
By the $s$-$m$-$n$ theorem, define a recursive function $g$ such that $\varphi_{g(e)}(x)$ is $1$, $0$, or undefined,
depending on whether $\varphi_e(x)$ is positive, zero, or undefined.
In particular, if $e$ is a characteristic index, then $g(e)$ is a characteristic index and $\varphi_{g(e)}=\varphi_e$.
Define $F$ on~$\mathcal{R}$ by $F(\varphi_e)=\varphi_{\delta''(g(e))}$.
Then $F$ is an effective operation on~$\mathcal{R}$ 
(depending on whether $g(e)$ is a characteristic index for a winning coalition or a losing coalition, 
$F(\varphi_e)=\varphi_{e_1}$ or $F(\varphi_e)=\varphi_{e_0}$).
By the theorem above, $F$ is effectively continuous on $\mathcal{R}$ so that for some recursive~$\psi$, 
(\ref{effectcont}) holds.  
Denote by $A_i$ the set $A_{\sigma}$ where $\sigma=\sigma(0)=i$.  
Since $F$ maps any $\varphi_e\in \mathcal{R}$ into constant functions
$\varphi_{e_1}\in A_1$ and $\varphi_{e_0}\in A_0$, we have $F^{-1}(\{\varphi_{e_i}\})=F^{-1}(A_i)$ 
for $i\in \{0,1\}$.
We therefore have $\varphi_e$ in $F^{-1}(A_1)$ or in $F^{-1}(A_0)$, depending on whether
$e$ is a characteristic index for a winning coalition or a losing coalition.
The $\Longrightarrow$ direction of Theorem~\ref{delta0det} is obtained by letting
$T_i$ be the r.e.\ set $\{t(a): a\in W_{\psi(t^{-1}(i))}\}$ restricted to the $0$-$1$ strings.

\subsection{Proof of Theorem~\ref{nakamura-thm} (Nakamura's theorem)}

($\Longleftarrow$).  Suppose that $X$ is finite, $\#X<\nu(\omega)$, and  $C(\omega, X, \p)=\emptyset$
for some measurable profile~$\p\in \profs$.
Then follow the proof of Theorem~2.5 in \citet{nakamura79} to find a cycle with respect to~$\succ^\p_\omega$
consisting of at most $\#X$ alternatives.

($\Longrightarrow$). 
Suppose $C(\omega, X, \p)\neq \emptyset$ for all~$\p\in \profs$.

(i)~To show that $X$ is finite, suppose it is infinite.  Then $X$ contains a countable subset
$X'=\{x_1, x_2, x_3, \ldots\}\subseteq X$.  Let $\p\in \A^{N'}$ be a profile such that 
all players $i\in N'$ have an identical preference $\pref$
(e.g., the transitive closure of itself) satisfying $x_{j+1}\pref x_j$ for all $j\in \{1, 2, \ldots\}$ and 
$x_1\pref y$ for all $y\in X\setminus X'$.
The measurability condition $\p\in\profs$ is satisfied since for all $x$, $y\in X$,
we have $\xprefy=N'$ or $\emptyset$, both in~$\B$.
Choose any winning coalition~$S\in\omega$, which exists by assumption.  
Then all players in~$S$ have the same preference~$\pref$, implying 
$x_{j+1}\succ^\p_\omega x_j$ for all $j$ and 
$x_1\succ^\p_\omega y$ for all $y\in X\setminus X'$.
It follows that  $C(\omega, X, \p)= \emptyset$; a contradiction.

(ii)~To show that $\#X<\nu(\omega)$, suppose $r:=\#X \geq \nu(\omega)$.  This excludes 
the possibility that $\omega$ is weak or $\nu(\omega)$ is infinite.  We will construct a profile~$\p$
such that the dominance relation~$\succ^\p_\omega$ has a cycle.
By the definition of the Nakamura number, there is a collection $\omega'=\{L_1, \ldots, L_r\}\subseteq \omega$
such that $\bigcap \omega'=\bigcap_{k=1}^r L_k = \emptyset$.  Define $L_0=N'$ and 
for all $k\in \{1, \ldots, r\}$, 
\[
D_k=(L_0\cap L_1\cap \cdots \cap L_{k-1})\setminus L_k.
\]
Then $\{D_1, \ldots, D_r\}$ is a family of (possibly empty) pairwise disjoint coalitions in~$\B$
such that $L_k\subseteq D_k^c:=N'\setminus D_k$ for all $k$ and $\bigcup_{k=1}^r  D_k=N'$
($i\in N'$ is in the first $D_k$ such that $i\notin L_k$).  
 
Write $X=\{x_1, \ldots, x_r\}$ and $x_0=x_r$.  Fix the cycle
\[
\succ\, =\{(x_k,x_{k-1}): k\in \{1, \ldots, r\} \}.
\]
Define $\p\in\A^{N'}$ as follows:  for each $k$, all players~$i$ in $D_k$ have the same (acyclic)
preference $\pref= \, \succ\setminus\{(x_k,x_{k-1})\}$.  Then for all $(x,y)\notin \, \succ$, we have
$\xprefy=\emptyset \in \B$.  On the other hand, for all $(x,y)=(x_k,x_{k-1}) \in \, \succ$, we have
$\xprefy=D_k^c\in \B$ and $L_k\subseteq D_k^c$.  
Therefore, $\p\in\profs$ and $\succ^\p_\omega=\, \succ$, a cycle.
It follows that $C(\omega, X, \p)=\emptyset$; a contradiction.

\subsection{Proof of Proposition \ref{filter-nakamura}}

(i)~If $\omega$ is a nonweak prefilter, then it has an infinite Nakamura number.
But nonweak computable games have a finite Nakamura number by Corollary~\ref{nakamura-finite}.

(ii) and (iii) are obvious from the following corollary \citep[Theorem 3.1]{nakamura79}
 of Theorem~\ref{nakamura-thm}:
$\succ^\p_\omega$ is acyclic for all $\p\in\profs$ if and only if $\#X'<\nu(\omega)$ for all finite $X'\subseteq X$. 
(This corollary can also be obtained from the well-known fact that $\succ^\p_\omega$
is acyclic if and only if the set $C(\omega,X',\p)$ of maximal elements with
respect to $\succ^\p_\omega$ is nonempty for all finite subsets $X'$ of $X$.)

\end{document}